\documentclass[aps,prd,reprint,letterpaper,groupedaddress,floatfix]{revtex4-2}
\usepackage{amsmath}
\usepackage{amssymb}
\usepackage{mathtools}
\usepackage{braket}
\usepackage{ascmac}
\usepackage{hyperref}
\usepackage{natbib}
\usepackage{graphicx}
\renewcommand{\v}[1]{\mbox{\boldmath $#1$}}
\begin{document}

%\preprint{}

\title{Quantum synchro-curvature masers and their application to neutron stars}


%\thanks{}
\author{Hiroko Tomoda}
\email[]{hrk-110-t@g.ecc.u-tokyo.ac.jp}
\affiliation{Department of Applied Physics, School of Engineering, The University of Tokyo, 7-3-1 Hongo, Bunkyo-ku, Tokyo 113-8656, Japan}

\author{Tomoya Naoe}
\noaffiliation

\author{Shoichi Yamada}
\email{shoichi@waseda.jp}
\altaffiliation{Advanced Research Institute for Science and Engineering, Waseda University, 3-4-1 Okubo, Shinjuku-ku, Tokyo 169-8555, Japan}
\affiliation{Department of Physics, School of Science and Engineering, Waseda University, 3-4-1 Okubo, Shinjuku-ku, Tokyo 169-8555, Japan}


\date{\today}

\begin{abstract}
  We explore the possibility of synchro-curvature maser in the magnetosphere of neutron stars (NSs).
  Unlike previous studies, we employ relativistic quantum mechanics, solving the Dirac equation for an electron in helical magnetic fields and calculating the radiative transition rates perturbatively.
  Assuming that the curvature of magnetic-field lines is much larger than the Larmor radius, we utilize adiabatic spinor rotations to obtain the wave functions of an electron.
  We classify the electron states further either by the spin operator projected on the magnetic field or by the helicity operator.
  We then evaluate numerically the true absorption rates accounting for the induced emission for some parameter values typical to the outer gaps of different types of NSs.
  We show that maser is indeed possible for a range of parameters.
  We will also present the dependence on those parameters systematically.
  We finally give a crude estimate of the amplification factor in the outer gap of NSs, which seems to favor millisecond pulsars as the host of maser emissions.
\end{abstract}

%\keywords{}

\maketitle


\section{Introduction\label{Introduction}}
Some astronomical objects are known to produce extreme radiations.
Fast radio bursts (FRBs) are one of such objects: they are radio transients known for their millisecond durations, large dispersion measures, and very high brightness temperatures.
They were discovered by Lorimer for the first time \cite{Lorimer2007} and were initially considered to be associated with some catastrophic events by many authors (see \cite{Lu2018, Cordes2019, Platts2019, Zhang2020, Xiao2021, Lyubarsky2021} for the catalogues of FRB mechanisms).
However, re-bursts from the same origin of FRB 121102 were observed in 2016 \cite{Spitler2016} and one-time events such as NS mergers and supernova explosions are disfavored at least as the repeating FRBs' mechanism.
Last year, some telescopes detected a radio burst from one of the Galactic magnetars, SGR 1935+2154 \cite{Andersen2020, Bochenek2020}.
Although no FRB had been observed in the Milky Way galaxy by that time, the energy was typical to FRBs and it was regarded as FRB, dubbed FRB 200428.

Whereas these observations have constrained the origin of FRB, the radiation mechanism is still largely unknown.
Many models have been advocated so far.
The very high brightness temperature, among other characteristics, requires a coherent mechanism.
Two types of coherent-radiation mechanisms are well-known: particle bunching and maser.
In the former case, charged particles are bunched with a size much smaller than the radiation wavelengths.
Then the emitted electromagnetic waves are positively superimposed on top of each other to produce high coherency.
The difficulty with this scenario is, of course, how to make such compact bunches.
In the latter case, on the other hand, coherent emission occurs as the induced emission.
In this paper we consider this maser emission from electrons and positrons moving in non-uniform magnetic fields.

In general, there are three types of radiations from charged particles in the magnetic field: synchrotron radiation, synchro-curvature radiation, and curvature radiation.
Synchrotron radiation is produced by a particle spiraling in a uniform magnetic field.
On the other hand, if the magnetic-field line is curved with a curvature radius much larger than the gyro-radius, the charged particle traces the same field line.
Then the acceleration induced by the curvature gives rise to the so-called curvature radiation.
In this case, the radiation looks a bit like the synchrotron radiation with a large Larmor radius.
Strictly speaking, however, the electron still has an angular momentum along the field line.
In the synchro-curvature radiation, both the gyration around the magnetic-field line and the translational motion along it are taken into account, and synchrotron radiation and curvature radiation are just two extremes \cite{Cheng1996, Kelner2015}.

Synchrotron maser was suggested for the radiation mechanism of FRB first in \cite{Lyubarsky2014} and then in \cite{Ghisellini2017} though magnetic-field lines emanating from astronomical objects such as pulsars are normally curved, which means that their curvature should be taken into account.
Curvature maser was once believed to be impossible \cite{Blandford1975}.
However, it turns out that it \textit{is} possible if one takes into account the drift from the magnetic-field line \cite{Zheleznyakov1979, Luo1992}.
Maser emissions were also shown to be possible if the field line has a torsion \cite{Luo1995, Luo1995}.
Synchro-curvature maser was discussed in \cite{Liu2007}.
Note that all these works are based on classical mechanics.

In quantum mechanics, the wave functions of a relativistic charged particle in the uniform magnetic field are well-known \cite{Sokolov1968} and were employed in the calculation of synchrotron radiation \cite{Latal1986, Harding1987}.
Recently, the wave functions of a charged particle in the circular magnetic field were derived and utilized for the quantum mechanical calculation of synchro-curvature radiation \cite{Voisin2017a, Voisin2017} though the maser possibility in synchro-curvature radiation has not been considered so far.

The purpose of this paper is to find the condition for the quantum synchro-curvature maser.
Following \cite{Voisin2017a}, we assume that the curvature of magnetic field is locally negligible.
We consider in this paper either a circular magnetic field as in \cite{Voisin2017a} or a helical one.
The paper is organized as follows.
In Sec.~\ref{Formulation}, we derive the wave functions of an electron in the helical magnetic field, solving the Dirac equation.
In so doing, we deploy the local helical coordinates and employ adiabatic spinor rotations.
Then we calculate the emission and absorption rates using the perturbation theory.
Finally, in Sec.~\ref{Results and Discussion}, we discuss possible parameter regions for the synchro-curvature maser and give a rough estimate of the amplification factor.

\section{Formulation\label{Formulation}}
In this section, we first write down the Dirac equation of a charged particle in a static magnetic field.
Then we solve it and derive the wave functions for a uniform magnetic field, which is not new.
Deploying the helical coordinates and applying adiabatic spinor rotations, we construct the wave functions for a helical field, which is original in this paper.
Finally, based on the wave functions so obtained, we formulate the emission and absorption rates of synchro-curvature radiation.

In this paper, we use CGS units unless otherwise stated.
We adopt the metric tensor given as
\begin{align}
  \eta^{\mu \nu}=
  \begin{pmatrix}
    1 & 0 & 0 & 0 \\
    0 & -1 & 0 & 0 \\
    0 & 0 & -1 & 0 \\
    0 & 0 & 0 & -1 \\
  \end{pmatrix}.
\end{align}
The electromagnetic potential and tensor in the Cartesian coordinates are defined as follows:
\begin{align}
  A^\mu&=\left(\phi,\v{A}\right),\\
  F^{\mu \nu}&=\partial^{\mu}A^{\nu}-\partial^{\nu}A^{\mu}
  =
  \begin{pmatrix}
    0 & -E^1 & -E^2 & -E^3\\
    E^1 & 0 &  -B^3 & B^2\\
    E^2 &  B^3 & 0 &  -B^1\\
    E^3 &  -B^2  & B^1 & 0\\
  \end{pmatrix},
\end{align}
where $\phi$ and $\v{A}$ are the scalar and vector potentials, respectively, and $\v{E}$ and $\v{B}$ are the electric and magnetic fields, respectively.
Latin and Greek letters run from 1 to 3 and from 0 to 3, respectively.
In the following, we employ Einstein's summation convention.

\subsection{Dirac equation of a charged particle in a static magnetic field\label{Dirac equation of a charged particle_in a static magnetic field}}
The Dirac equation of a particle with a charge $q$ is given as
\begin{gather}
  \left(i\hbar \gamma^\mu D_\mu-mc\right)\Phi(\v{r},t)=0,\label{D1}\\
  D_\mu\equiv\partial_\mu+i\frac{q}{\hbar c}A_\mu(\v{r},t).
\end{gather}
Defining the momentum operator as
\begin{subequations}
  \begin{align}
    \pi_0&\equiv -i\hbar D_0,\\
    \pi_i&\equiv -i\hbar D_i=-i\hbar \nabla-\frac{q\v{A}(\v{r})}{c}\equiv \v{\pi},
  \end{align}
\end{subequations}
and multiplying Eq.~\eqref{D1} with $\gamma^0$ from the left, we get
\begin{align}
  &\left(\gamma^0\gamma^0 \pi_0 + \gamma^0\gamma^i \pi_i+mc \gamma^0 \right)\Phi(\v{r},t)=0.\label{D1_1}
\end{align}
Since we consider only static magnetic fields in this paper, the wave function and the zeroth component of the momentum operator can be written as
\begin{gather}
  \Phi(\v{r},t)
  =e^{-i\frac{E}{\hbar}t}\Psi(\v{r})
  ,\\
  \pi_0=-\frac{E}{c},
\end{gather}
where $E$ is particle's energy.
In the following, we will omit the negative energy solution as we are not interested in pair processes.

We choose the Weyl representation, in which the gamma matrices are written as follows:
\begin{gather}
  \begin{gathered}
    \gamma^0=
    \begin{pmatrix}
      0 & I_2\\
      I_2 & 0\\
    \end{pmatrix}
    ,
    \gamma^i=
    \begin{pmatrix}
      0 & -\sigma^i\\
      \sigma^i & 0\\
    \end{pmatrix}
    ,
    \gamma^5
    =\begin{pmatrix}
      I_2 & 0\\
      0 & -I_2\\
    \end{pmatrix}
    ,
  \end{gathered}
  \\
  \sigma^1=
  \begin{pmatrix}
    0 & 1\\
    1 & 0\\
  \end{pmatrix}
  ,
  \sigma^2=
  \begin{pmatrix}
    0 & -i\\
    i & 0\\
  \end{pmatrix}
  ,
  \sigma^3=
  \begin{pmatrix}
    1 & 0\\
    0 & -1\\
  \end{pmatrix}.
\end{gather}
In the above expressions, $\sigma^i (i = 1 \sim 3)$ are the Pauli matrices and $I_2$ is the $2\times 2$ identity matrix.
Introducing two-component spinors as $\Psi(\v{r})=\left(\varphi_R(\v{r}),\varphi_L(\v{r})\right)$, we write Eq.~\eqref{D1_1} as
\begin{gather}
  \left[\pi_0 +
  \begin{pmatrix}
    \sigma^i & 0\\
    0 & -\sigma^i\\
  \end{pmatrix}
  \pi_i+mc
  \begin{pmatrix}
    0 & I_2\\
    I_2 & 0\\
  \end{pmatrix}
  \right]
  \begin{pmatrix}
    \varphi_R(\v{r})\\
    \varphi_L(\v{r})\\
  \end{pmatrix}
  =0.\label{D2}
\end{gather}
Eliminating $\varphi_L(\v{r})$ in Eq.~\eqref{D2}, we obtain the equation for $\varphi_R(\v{r})$ as
\begin{align}
  &\left[\pi_0^2+\sigma^i\left(\pi_0 \pi_i-\pi_i \pi_0\right)-\left(\sigma^i \pi_i \right)^2
  \right.\nonumber \\
  &\hspace{40mm}
  -m^2c^2\Bigr]\varphi_R(\v{r})=0.
\end{align}
The third term on the left hand side of the above equation is re-arranged as follows:
\begin{subequations}
  \begin{align}
    \left(\sigma^i \pi_i \right)^2
    &=\left(\sigma^i \pi_i \right)\left(\sigma^j \pi_j \right)\nonumber \\
    &=\left(\delta ^{ij}+i\varepsilon^{ijk} \sigma_k \right)\pi_i \pi_j\nonumber \\
    &=\v{\pi}^2+i\v{\sigma}\cdot \left(\v{\pi}\times \v{\pi}\right),
    \\
    \v{\pi}\times \v{\pi}
    &=\left(-i\hbar \nabla-\frac{q\v{A}(\v{r})}{c}\right)
    \times
    \left(-i\hbar \nabla-\frac{q\v{A}(\v{r})}{c}\right)\nonumber \\
    &=i \frac{q \hbar}{c} \left[\v{A}(\v{r})\times \nabla +\left(\nabla \times \v{A}(\v{r})\right)-\v{A}(\v{r})\times \nabla \right]\nonumber \\
    &=i \frac{q \hbar}{c} \v{B}(\v{r}),
  \end{align}
\end{subequations}
where $\delta^{ij}$ is the Kronecker delta, $\varepsilon^{ijk}$ is the Levi-Civita symbol and $\v{\sigma}=(\sigma^1,\sigma^2,\sigma^3)$.
Then the 2-component wave functions of a charged particle in a static magnetic field satisfy the following equations:
\begin{gather}
  \left[\left(-i\hbar \nabla-\frac{q\v{A}(\v{r})}{c} \right)^2-\frac{q\hbar}{c} \v{\sigma} \cdot \v{B}(\v{r}) \right]\varphi_{R,L}(\v{r}) \nonumber\\
  \hspace{35mm}=\frac{E^2-m^2c^4}{c^2}\varphi_{R,L}(\v{r}),\label{D3}
  \\
  \begin{cases}
    \left(E-c\v{\sigma}\cdot\v{\pi} \right) \varphi_R(\v{r})=mc^2\varphi_L(\v{r})\\
    \left(E+c\v{\sigma}\cdot\v{\pi} \right) \varphi_L(\v{r})=mc^2\varphi_R(\v{r})
  \end{cases}.\label{D4}
\end{gather}
Note that they are valid in the Cartesian coordinates.

\subsection{Uniform magnetic field\label{Uniform magnetic field}}
We first consider the uniform magnetic field parallel to the $z$-axis.
Although the wave functions are well-known in this case, we explain them because we will employ them for the helical field later.

\subsubsection{Wave functions of an electron\label{Wave functions of an electron}}
In the cylindrical coordinates, the vector potential for the magnetic field of our current concern is given in the Coulomb gauge as
\begin{align}
  \v{A}&=\frac{1}{2}Br\v{e_\phi}
\end{align}
and the corresponding contravariant and covariant four-vectors are written as
\begin{subequations}
  \begin{align}
    A^\mu&=\left(A^0,A^r,A^\phi,A^z\right)=\left(0,0,\frac{1}{2}B,0\right),\\
    A_\mu&=\left(A_0,A_r,A_\phi,A_z\right)=\left(0,0,-\frac{1}{2}Br^2,0\right).
  \end{align}
\end{subequations}
In this gauge, the first term on the left hand side of Eq.~\eqref{D3} is rewritten as
\begin{align}
  &\left(-i\hbar \nabla+\frac{e\v{A}(\v{r})}{c} \right)^2\varphi_{R,L}(\v{r})\nonumber \\
  &=\left(-\hbar^2 \nabla^2
  -2i\frac{e\hbar}{c} \v{A}(\v{r}) \cdot \nabla
  +\frac{e^2}{c^2}\v{A}(\v{r})^2\right)\varphi_{R,L}(\v{r})\label{D4_1},
\end{align}
where $e$ is the elementary charge (positive quantity).
Using the Larmor radius $\lambda=\sqrt{2\hbar c/eB}$, we normalize the independent variables as $z=\lambda u,\ r=\lambda v$.
Then Eq.~\eqref{D3} is re-arranged into
\begin{align}
  &\left(\partial_v ^2+\frac{1}{v}\partial_v
  +\frac{1}{v^2}\partial_\phi ^2+\partial_u ^2+2i\partial_\phi
  -v^2-2\sigma^3 \right) \varphi_{R,L}(\v{r})\nonumber\\
  &\hspace{20mm}=-\left(\frac{\lambda}{\hbar c}\right)^2\left(E^2-m^2c^4\right)\varphi_{R,L}(\v{r}).\label{D5}
\end{align}

We express $\varphi_{R} (\v{r})$ in the following form:
\begin{gather}
  \varphi_{R} (\v{r})=
  \begin{pmatrix}
    e^{i l_{\perp +} \phi}e^{i l_{\parallel +} u}\xi_+(v)\\
    e^{i l_{\perp -} \phi}e^{i l_{\parallel -} u}\xi_-(v)\\
  \end{pmatrix},
\end{gather}
where $l_{\perp \pm}$ are integers because of the boundary condition and $l_{\parallel \pm}$ are real numbers.
Hereafter we solve Eq.~\eqref{D5} for $\varphi_R(\v{r})$.
The equations of $\xi_\pm(v)$ are obtained as
\begin{align}
  \left(\partial_v ^2+\frac{1}{v}\partial_v-\frac{l_{\perp \pm}^2}{v^2}-l_{\parallel \pm} ^2-2l_{\perp \pm} -v^2\mp 2\right)\xi_\pm(v)\nonumber\\
  =-\left(\frac{\lambda}{\hbar c}\right)^2\left(E^2-m^2c^4\right)\xi_\pm(v),
\end{align}
or, more conveniently, as
\begin{align}
  \left[\partial_v ^2+\frac{\frac{1}{4}-l_{\perp \pm}^2}{v^2}-v^2-l_{\parallel \pm} ^2-2l_{\perp \pm} \mp 2
  \right.\hspace{15mm}\nonumber
  \\
  +
  \left.
  \left(\frac{\lambda}{\hbar c}\right)^2\left(E^2-m^2c^4\right)\right]\left(\sqrt{v}\xi_\pm(v)\right)=0.\label{D6}
\end{align}
The above equation should be compared with the following one:
\begin{gather}
  \begin{gathered}
    \left(\partial_x ^2+\frac{\frac{1}{4}-{\alpha}^2}{x^2}-x^2+4s+2\alpha+2\right)h(x)=0,\\
    \hspace{35mm}(\alpha >-1,s=0,1,2 \cdots)
  \end{gathered} \label{D6_1}
\end{gather}
for which we know the solutions \cite[Table~18.8.1]{NIST:DLMF} as
\begin{gather}
  h(x)=e^{-\frac{x^2}{2}}x^{\alpha+\frac{1}{2}}L_s^{\left(\alpha\right)} \left(x^2 \right).\label{D6_1-1}
\end{gather}
Here $L_s^{\left(\alpha\right)}(x)$ is the Laguerre polynomial defined as
\begin{align}
  L_s ^{(\alpha)} (x) \equiv x^{-\alpha} e^{x} \frac{d^s}{d x^s}
  \left(x^{\alpha+s} e^{-x}\right)
  \quad (s,\alpha=0,1,2\cdots).\label{Laguerre}
\end{align}

With this formula, we obtain $\xi_{\pm}(v)$ as the case of $\alpha=l_{\perp}\geq 0$ as follows:
\begin{gather}
  \xi_{\pm} (v)=C_{R_\pm} e^{-\frac{v^2}{2}}v^{l_{\perp \pm}}L_{s_\pm}^{\left(l_{\perp \pm} \right)} \left(v^2 \right),
  \\
  s_\pm=\frac{1}{4}\left[-l_{\parallel \pm} ^2-4 l_{\perp \pm} \mp2 -2+\left(\frac{\lambda}{\hbar c}\right)^2\left(E^2-m^2c^4\right)\right], \nonumber
\end{gather}
where $C_{R_+},C_{R_-}$ are arbitrary constants.
Although one may also take $\alpha=-l_{\perp}>0$, the resultant solution is equivalent to the above one (see Appendix \ref{B} for details). We hence use the above functions alone.
Of the two possible relations between $l_{\perp \pm}$ and $s_\pm$:
\begin{align}
  \begin{dcases}
    l_{\perp +}=l_{\perp -} -1 & \text{and  } s_+=s_-,\\
    l_{\perp +}=l_{\perp -} & \text{and  } s_+=s_- -1,
  \end{dcases}\label{D7}
\end{align}
only the former is compatible with Eq.~\eqref{D4}. We rewrite them accordingly as follows:
\begin{subequations}
  \begin{align}
    l_{\parallel +}=l_{\parallel -}&\equiv l_{\parallel},
    \\
    l_{\perp +}+1=l_{\perp -}&\equiv l_{\perp},
    \\
    s_+=s_-&\equiv s,
    \\
    s+l_{\perp}&\equiv n.\label{D7-0}
  \end{align}
\end{subequations}
It is noted that $l_{\parallel}$ is a real number, whereas $n$ and $s$ are non-negative integers, and $l_{\perp}$ is an integer satisfying $l_{\perp}\leq n$.
Finally, the solution of Eq.~\eqref{D5} is given as
\begin{gather}
  \varphi_R(\v{r})=
  \begin{pmatrix}
    C_{R_+} e^{i (l_\perp-1) \phi}e^{i l_\parallel u}
    e^{-\frac{v^2}{2}}v^{l_\perp-1}L_s^{\left(l_\perp -1\right)}  \left(v^2 \right)\\
    C_{R_-} e^{i l_{\perp} \phi}e^{i l_\parallel u}
    e^{-\frac{v^2}{2}}v^{l_\perp}L_s^{\left(l_\perp \right)}  \left(v^2 \right)\\
  \end{pmatrix}\label{D7-1}
  ,
  \\
  E_{n,l_\parallel}=\sqrt{m^2 c^4 +\left(\frac{\hbar c}{\lambda}\right)^2\left(4n +l_\parallel ^2 \right)}
  .\label{D7-2}
\end{gather}
The other two-component spinor $\varphi_L(\v{r})$ can be obtained in the similar way as
\begin{align}
  \varphi_L(\v{r})=
  \begin{pmatrix}
    C_{L_+} e^{i (l_\perp-1) \phi}e^{i l_\parallel u}
    e^{-\frac{v^2}{2}}v^{l_\perp-1}L_s^{\left(l_\perp -1\right)}  \left(v^2 \right)\\
    C_{L_-} e^{i l_{\perp} \phi}e^{i l_\parallel u}
    e^{-\frac{v^2}{2}}v^{l_\perp}L_s^{\left(l_\perp \right)}  \left(v^2 \right)
  \end{pmatrix}
  ,
\end{align}
where $C_{L_+},C_{L_-}$ are arbitrary constants.

Since the electron is massive, the right-handed and left-handed spinors are coupled with each other and the four integral constants are not completely arbitrary.
In fact, inserting Eqs.~\eqref{D7-1}, \eqref{D7-2} into Eq.~\eqref{D4}, we get the following relations:
\begin{gather}
  \begin{dcases}
    C_{R_-}=
    \frac{1}{2ni}\frac{\lambda}{\hbar c}\left[
     -\left(E_{n,l_\parallel}-\frac{\hbar c l_\parallel}{\lambda}\right)C_{R_+}+mc^2 C_{L_+}
     \right]
     \\
    C_{L_-}=
    \frac{1}{2ni}\frac{\lambda}{\hbar c}\left[
    -mc^2 C_{R_+}+\left(E_{n,l_\parallel}+\frac{\hbar c l_\parallel}{\lambda}\right)C_{L_+}
    \right]
  \end{dcases}
  .
  \label{D8}
\end{gather}
In so doing, we utilized the following formulae:
\begin{subequations}
  \begin{align}
    &\frac{d}{dx} L_s^{(\alpha)}\left(x^2\right)
    =-2x L_{s-1}^{(\alpha+1)}\left(x^2\right),
    \\
    &L_s^{(\alpha)}\left(x^2\right)
    +L_{s-1}^{(\alpha+1)}\left(x^2\right)
    =L_s^{(\alpha+1)}\left(x^2\right),
    \\
    &(s+\alpha+1)L_s^{(\alpha)}\left(x^2\right)\hspace{40mm}\nonumber \\
    &\hspace{10mm}=
    \left(\alpha+1\right)L_s^{(\alpha+1)}\left(x^2\right)-x^2L_{s-1}^{(\alpha+2)}\left(x^2\right).
  \end{align}
\end{subequations}

The eigenvalue, or the energy, is a function of quantum numbers $n$ and $l_{\parallel}$:
\begin{gather}
  E_{n,l_\parallel}=\sqrt{m^2 c^4 +2m^2 c^4\frac{B}{B_c}n +\left(c P_z\right)^2},\label{D9_1}
  \\
  B_c\equiv \frac{m^2c^3}{e\hbar}=4.414\times10^{13} \text{G},
  \\
  P_z\equiv \frac{\hbar l_\parallel}{\lambda}.
\end{gather}
In Eq.~\eqref{D9_1}, the terms in the square root represent from the left the rest mass energy, the energy associated with the quantized cyclotron motion, to which we refer as the Landau level, and the energy of the translational motion along the field line;
$B_c$ is the critical field strength, at which the Landau level becomes comparable to the rest mass energy.
To each value of energy, there is a degeneracy with respect to $s$, which is corresponding to the degree of freedom in the position of the guiding center for cyclotron motion.
The eigenstate for a given trio of quantum numbers ($n$, $s$, $l_{\parallel}$) is written as
\begin{gather}
  \Psi_{n,s,l_{\parallel}}(\v{r})
  =
  \begin{pmatrix}
    C_{R_+} e^{i (l_\perp-1) \phi}e^{i l_\parallel u}
    e^{-\frac{v^2}{2}}v^{l_\perp-1}L_s^{\left(l_\perp -1\right)}  \left(v^2 \right)\\
    C_{R_-} e^{i l_{\perp} \phi}e^{i l_\parallel u}
    e^{-\frac{v^2}{2}}v^{l_\perp}L_s^{\left(l_\perp \right)}  \left(v^2 \right)\\
    C_{L_+} e^{i (l_\perp-1) \phi}e^{i l_\parallel u}
    e^{-\frac{v^2}{2}}v^{l_\perp-1}L_s^{\left(l_\perp -1\right)}  \left(v^2 \right)\\
    C_{L_-} e^{i l_{\perp} \phi}e^{i l_\parallel u}
    e^{-\frac{v^2}{2}}v^{l_\perp}L_s^{\left(l_\perp \right)}  \left(v^2 \right)
  \end{pmatrix}
  ,\label{D9}
\end{gather}
where Eq.~\eqref{D7-0} gives $l_{\perp}$ in terms of $n$ and $s$ as
\begin{gather}
  l_{\perp}= n-s.
\end{gather}
In Eq.~\eqref{D9}, the normalization is not considered yet and $C_{R_+}$ and $C_{L_+}$ are arbitrary constants (the other coefficients $C_{R_-}$ and $C_{L_-}$ are determined from Eq.~\eqref{D8}).
% ; the range of $s$ is determined from Eq.~\eqref{D6_1}.

\subsubsection{Helicity and spin eigenstates\label{Helicity and spin eigenstates}}
The degenerate eigenstates obtained just now can be put in more physically motivated forms.
Using the normalized function:
\begin{gather}
  f_{l_{\perp}}(\v{r})
  =\frac{1}{\sqrt{L}}e^{il_\parallel u}
  \frac{1}{\sqrt{2\pi}}e^{i l_{\perp}\phi}
  \frac{1}{\lambda}\sqrt{\frac{2\cdot s!}{n!}}
  e^{-\frac{v^2}{2}}v^{l_{\perp}}L_s^{(l_\perp)}\left(v^2\right)
  ,\label{f_l}
\end{gather}
the wave function $\Psi_{n,s,l_{\parallel}}(\v{r})$ can be written as
\begin{gather}
  \Psi_{n,s,l_{\parallel}}(\v{r})
  =
  \begin{pmatrix}
    c_1 f_{l_{\perp}-1}(\v{r})\\
    c_2 f_{l_{\perp}}(\v{r})\\
    c_3 f_{l_{\perp}-1}(\v{r})\\
    c_4 f_{l_{\perp}}(\v{r})\\
  \end{pmatrix}
  ,
\end{gather}
where coefficients $c_1\sim c_4$ are constants to be determined, depending on the eigenstates; $L$ is the quantization length along the magnetic-field line.
Note that the subscripts for other quantum numbers $n,s(=n-l_{\perp}),l_{\parallel}$ are dropped for notational simplicity in Eq.~\eqref{f_l}.

There are some types of the spin operator (see \cite{Bordovitsyn1999, Melrose1983b} for the review).
We here use the one defined in \cite{Bargmann1948, Fradkin1961}, which is called the \textit{vector operator of transverse polarization} in \cite[Eqs.~(2.28), (2.29)]{Bordovitsyn1999}, as
\begin{gather}
  \Sigma_\mu
  \equiv
  \frac{\hbar}{2} \left(-\gamma_\mu+\frac{1}{mc}\pi_\mu\right)\gamma^5.
\end{gather}
Its $z$-component is expressed in the Weyl representation as
\begin{gather}
  \Sigma_z
  =\frac{\hbar}{2}
  \begin{pmatrix}
    \frac{\pi_z}{mc}  & \sigma^3\\
    \sigma^3 & -\frac{\pi_z}{mc} \\
  \end{pmatrix},
\end{gather}
which commutes with the Hamiltonian operator.
We consider the covariant operator in accordance with \cite[section 3.3.2]{Bordovitsyn1999}.
Then the energy eigenstates can be chosen as the eigenstates of this spin operator.
This is accomplished by choosing the coefficients as
\begin{gather}
  \begin{dcases}
    c_1=
    \frac{1}{2}\left(1\pm K_1\right)^{\frac{1}{2}}
    \left(1\pm K_2\right)^{\frac{1}{2}}
    \\
    c_2=
    \frac{i}{2}\left(1\mp K_1\right)^{\frac{1}{2}}
    \left(1\pm K_2\right)^{\frac{1}{2}}
    \\
    c_3=
    \pm \frac{1}{2}
    \left(1\pm K_1\right)^{\frac{1}{2}}
    \left(1\mp K_2\right)^{\frac{1}{2}}
    \\
    c_4=
    \mp \frac{i}{2}
    \left(1\mp K_1\right)^\frac{1}{2}
    \left(1\mp K_2\right)^{\frac{1}{2}}
  \end{dcases},
\end{gather}
\begin{subequations}
  \begin{gather}
    K_1 \equiv \frac{\sqrt{m^2c^4+\left(cP_z\right)^2}}{E}, \label{K1}
    \\
    K_2 \equiv \frac{cP_z}{\sqrt{m^2c^4+\left(cP_z\right)^2}}. \label{K2}
  \end{gather}
\end{subequations}
The upper and lower signs in the above equations are corresponding to the following eigenvalues of the spin operator:
\begin{align}
  \Sigma_z=\pm \frac{\hbar}{2}\frac{\sqrt{m^2 c^4 + (cP_z)^2}}{mc^2}.
\end{align}
% \cite{Melrose1983a, Bordovitsyn1999, Bargmann1948}

Next, we consider the helicity states.
They are the eigenstates of the helicity operator defined as
\begin{align}
  \hat{h}=\frac{\hbar}{2}\frac{1}{|\v{\pi}|}
  \begin{pmatrix}
    \v{\sigma}\cdot\v{\pi} & 0 \\
    0 & \v{\sigma}\cdot\v{\pi} \\
  \end{pmatrix}
  .
\end{align}
This operator commutes with the Hamiltonian for a static magnetic field.
The coefficients for the helicity eigenstates are given as follows:
\begin{gather}
  \begin{dcases}
    c_1=
    \frac{1}{2}
    \left(1\pm K_3\right)^{\frac{1}{2}}
    \left(1\pm K_4\right)^{\frac{1}{2}}
    \\
    c_2=
    \pm
    \frac{i}{2}
    \left(1\pm K_3\right)^{\frac{1}{2}}
    \left(1\mp K_4\right)^{\frac{1}{2}}
    \\
    c_3=
    \frac{1}{2}
    \left(1\mp K_3\right)^{\frac{1}{2}}
    \left(1\pm K_4\right)^{\frac{1}{2}}
    \\
    c_4=
    \pm
    \frac{i}{2}
    \left(1\mp K_3\right)^{\frac{1}{2}}
    \left(1\mp K_4\right)^{\frac{1}{2}}
  \end{dcases},
\end{gather}
\begin{subequations}
  \begin{gather}
    K_3 \equiv \frac{\sqrt{E^2-m^2c^4}}{E},\\
    K_4 \equiv \frac{cP_z}{\sqrt{E^2-m^2c^4}}.
  \end{gather}
\end{subequations}
The corresponding helicity eigenvalues are $\hat{h}=+\hbar/2$ for the upper case and $\hat{h}=-\hbar/2$ for the lower case.
In calculating the radiative transition rates in Sec.~\ref{Synchro-curvature maser}, we will employ these eigenstates and, in so doing, these coefficients will be used.

\subsection{Helical magnetic field\label{Formulation_Helical magnetic field}}
\begin{figure}[!hbtp]
  \includegraphics[width=80mm]{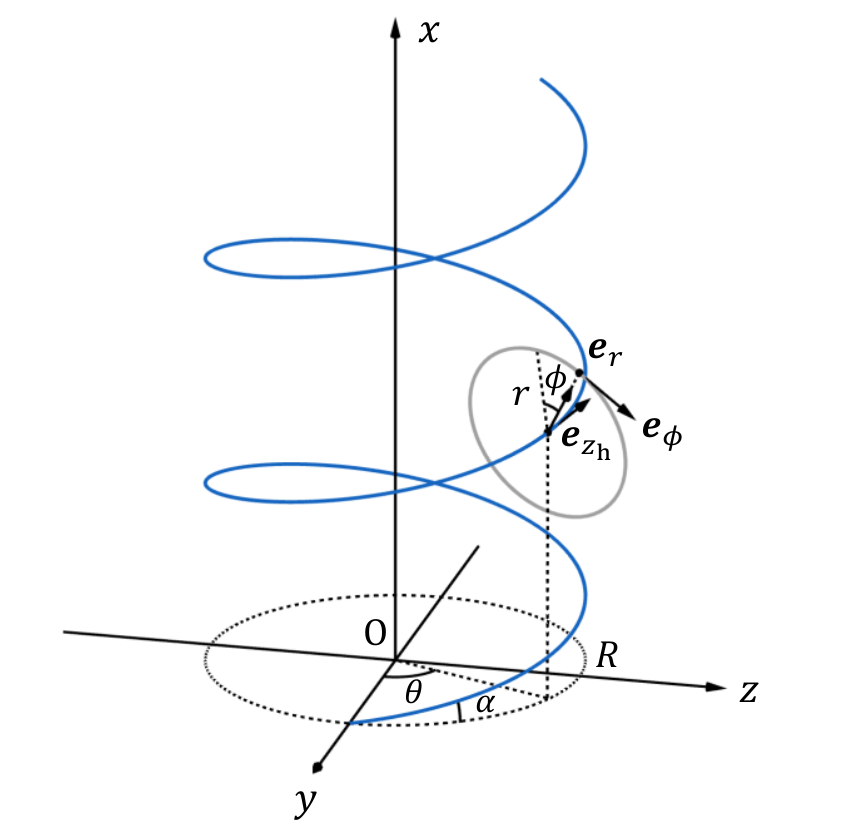}
  \caption{\label{Helical}Helical magnetic field}
\end{figure}

Now the original contents.
We consider a helical magnetic field like Fig.~\ref{Helical}.
We will first derive the wave functions of a charged particle moving in this field and then use the results for the study of quantum-synchro curvature maser.

In Fig.~\ref{Helical}, $\alpha$ is the pitch angle and $\alpha=0$ corresponds to a circular magnetic field \cite{Voisin2017a,Voisin2017}.
We deploy the helical coordinates (see Appendix \ref{A}), which are related with the Cartesian coordinates as
\begin{gather}
  \begin{dcases}
    x=z_{\text{h}} \sin{\alpha}+r\cos{\phi}\cos{\alpha}
    \\
    y=(R+r\sin{\phi})\cos{\theta}
    +r\cos{\phi}\sin{\alpha}\sin{\theta}
    \\
    z=(R+r\sin{\phi})\sin{\theta}
    -r\cos{\phi}\sin{\alpha}\sin{\theta}
  \end{dcases},
  \\
  \theta=\frac{z_{\text{h}} \cos{\alpha}}{R}.
\end{gather}
In these equations, $z_{\text{h}}$ is the length along the helix (see Appendix \ref{A} for details).
From now on, we assume that the curvature radius of the magnetic-field line is much larger than the Larmor radius.
In the outer gap of a NS, for example, the typical field strength is $B\sim 10^2$G, for which the Larmor radius is $\lambda \sim 10^{-5}$cm whereas the curvature radius is $R\sim 10^9$cm.
The assumption is also compatible with the local nature of the helical coordinates.
With this assumption, the 33-componetnt of Jacobian matrix for the helical coordinates, Eq.~\eqref{Jacobian}, is approximated as
\begin{align}
  g_{33}&=\sin^2{\alpha}+\frac{\cos^2{\alpha}}{R^2}\left(R + r\sin{\phi}\right)^2 \nonumber\\
  &\quad+\frac{r^2}{R^2}\cos^2{\alpha}\sin^2{\alpha}\cos^2{\phi}
  \approx 1.
\end{align}
Then the divergence and rotation of arbitrary scalar- and vector-valued functions, $f(\v{r})$, $\v{V}(\v{r})$, are given respectively as
\begin{align}
  \nabla f(\v{r})&=\v{e}_r \frac{1}{\sqrt{g_{11}}}\partial_r f
  +\v{e}_{\phi} \frac{1}{\sqrt{g_{22}}}\partial_{\phi} f
  +\v{e}_{z_{\text{h}}} \frac{1}{\sqrt{g_{33}}}\partial_{z_{\text{h}}} f \nonumber\\
  &\approx
  \v{e}_r\partial_r f
  +\v{e}_{\phi} \frac{1}{r}\partial_{\phi} f
  +\v{e}_{z_{\text{h}}} \partial_{z_{\text{h}}} f,
  \label{D9-1}
  \displaybreak[1] \\
  \nabla \times \v{V}(\v{r})
  &=\frac{1}{\sqrt{g_{11}g_{22}g_{33}}}
  \begin{vmatrix}
    \sqrt{g_{11}} \v{e}_r & \sqrt{g_{22}} \v{e}_{\phi} & \sqrt{g_{33}} \v{e}_{z_{\text{h}}} \\
    \partial_r & \partial_{\phi} & \partial_{z_{\text{h}}}\\
    \sqrt{g_{11}} V_r & \sqrt{g_{22}} V_{\phi} & \sqrt{g_{33}} V_{z_{\text{h}}} \\
  \end{vmatrix}
  \nonumber
  \\
  &\approx
  \left(\frac{1}{r}\partial_\phi V_{z_{\text{h}}} -\partial_{z_{\text{h}}} V_\phi \right)\v{e}_r
  +\left(\partial_{z_{\text{h}}} V_r- \partial_r V_{z_{\text{h}}}\right)\v{e}_\phi\nonumber\\
  &\quad +\frac{1}{r}\left(\partial_r\left(rV_\phi\right)-\partial_\phi V_r\right)\v{e}_{z_{\text{h}}}.
  \label{D9-2}
\end{align}

Note the similarity of Eqs.~\eqref{D9-1} and \eqref{D9-2} to the counterparts for the cylindrical coordinates.
In fact, the helical coordinates can be interpreted as local cylindrical coordinates.
This also means that the helical magnetic field can be regarded as a uniform field locally on these coordinates.
Then the vector potential may be written as
\begin{align}
  \v{A}&=\frac{1}{2}Br \v{e_\phi}
\end{align}
in the Coulomb gauge and the corresponding contravariant and covariant four-vectors are given as
\begin{subequations}
  \begin{align}
    A^{\mu}&=\left(A^0, A^r, A^\phi, A^{z_{\text{h}}} \right)=\left(0,0, \frac{1}{2}B, 0\right),\\
    A_{\mu}&=\left(A_0, A_r, A_\phi, A_{z_{\text{h}}} \right)=\left(0,0, -\frac{1}{2}Br^2, 0\right).
  \end{align}
\end{subequations}
\begin{widetext}
  \noindent
  Normalizing $r$ and $z_{\text{h}}$ with the Larmor radius: $z_{\text{h}}=\lambda u,\ r=\lambda v$, we recast Eq.~\eqref{D3} as
  \begin{align}
    \begin{aligned}
      \left[\partial_v ^2+\frac{1}{v}\partial_v+\frac{1}{v^2}\partial_\phi ^2+\partial_u ^2+2i\partial_\phi -v^2-2
        \begin{pmatrix}
          \cos{\alpha}\cos{\theta} & \sin{\alpha}+i\cos{\alpha}\sin{\theta}\\
          \sin{\alpha}-i\cos{\alpha}\sin{\theta} & -\cos{\alpha}\cos{\theta}\\
        \end{pmatrix}
      \right] \varphi_R(\v{r})\\
      =-\left(\frac{\lambda}{\hbar c}\right)^2\left(E^2-m^2c^4\right)\varphi_R(\v{r}),
    \end{aligned}\label{D10}
  \end{align}
which should be compared with the equation for the globally uniform magnetic field:
  \begin{align}
    \left[\partial_v ^2+\frac{1}{v}\partial_v+\frac{1}{v^2}\partial_\phi ^2+\partial_u ^2+2i\partial_\phi -v^2-2
      \begin{pmatrix}
        1 & 0\\
        0 & -1\\
      \end{pmatrix}
     \right] \varphi_R(\v{r})
    =-\left(\frac{\lambda}{\hbar c}\right)^2\left(E^2-m^2c^4\right)\varphi_R(\v{r}).\label{D10-1}
  \end{align}
\end{widetext}
They are indeed the same except for the Hermitian matrices in the last terms on the left hand side.
They are originated from the coupling between the electron's spin and the magnetic field.

\begin{figure}[!hbtp]
  \includegraphics[width=86mm]{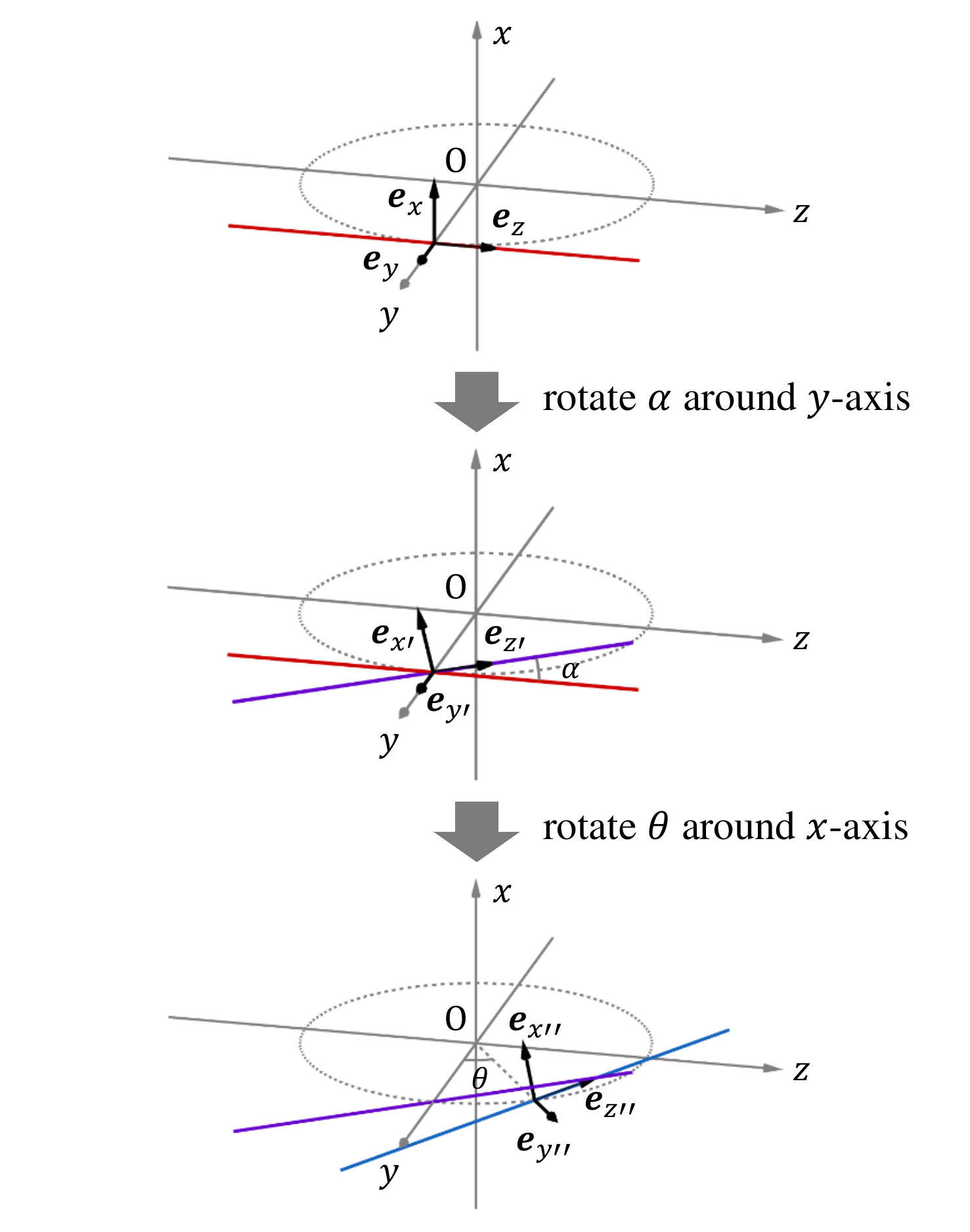}
  \caption{\label{Spinor_rotation}Frame rotations}
\end{figure}

The matrix in Eq.~\eqref{D10} can be diagonalized by an appropriate spinor rotation to the matrix in Eq.~\eqref{D10-1}.
Since this rotation does not commute with $\partial_u^2$ (note the rotation depends on $\theta$ and hence on $u$), Eq.~\eqref{D10} cannot be transformed to Eq.~\eqref{D10-1} in general.
Under the current approximation, however, the non-commutativity is negligibly small: the spinor rotation changes with $u$ very slowly on the scale of the curvature of the magnetic-field line, which is much larger than the Larmor radius.
In this adiabatic limit, we can solve Eq.~\eqref{D10} just by the diagonalization.
Defining $\Psi_{\text{hel}}(\v{r})$, $\Psi_{\text{uni}}(\v{r})$ as the wave functions in the helical and uniform magnetic fields, respectively, we obtain the followings:
\begin{gather}
  \Psi_{\text{hel}}(\v{r})
  =S^{23}(\theta) S^{31}(\alpha)
  \Psi_{\text{uni}}(\v{r}),
  \\
  S^{31}(\alpha)\equiv
  \begin{pmatrix}
    \cos{\frac{\alpha}{2}} & -\sin{\frac{\alpha}{2}} & 0 & 0\\
    \sin{\frac{\alpha}{2}} & \cos{\frac{\alpha}{2}} & 0 & 0\\
    0 & 0 & \cos{\frac{\alpha}{2}} & -\sin{\frac{\alpha}{2}}\\
    0 & 0 & \sin{\frac{\alpha}{2}} & \cos{\frac{\alpha}{2}}\\
  \end{pmatrix},
  \\
  S^{23}(\theta)\equiv
  \begin{pmatrix}
    \cos{\frac{\theta}{2}} & -i\sin{\frac{\theta}{2}} & 0 & 0\\
    -i\sin{\frac{\theta}{2}} & \cos{\frac{\theta}{2}} & 0 & 0\\
    0 & 0 & \cos{\frac{\theta}{2}} & -i\sin{\frac{\theta}{2}}\\
    0 & 0 & -i\sin{\frac{\theta}{2}} & \cos{\frac{\theta}{2}}\\
  \end{pmatrix},
  \displaybreak[1] \\
  E_{n,l_\parallel}=\sqrt{m^2 c^4 +2m^2 c^4\frac{B}{B_c}n +\left(c P_{z_{\text{h}}}\right)^2}
  ,\\
  P_{z_{\text{h}}} \equiv \frac{\hbar l_\parallel}{\lambda},
\end{gather}
where $S^{31}(\alpha), S^{23}(\theta)$ represent the spinor rotations involved, i.e., the rotation by the angle of $\alpha$ around the $y$-axis and the rotation by the angle of $\theta$ around the $x$-axis, respectively.
These rotations, combined, transform the basis vectors in the cylindrical coordinates to coincide with those in the helical coordinates (see Fig.~\ref{Spinor_rotation}).
Note that the eigenstates of the spin operator and the helicity operator are also rotated with $S^{23}(\theta),S^{31}(\alpha)$ in the same way.

\subsection{Synchro-curvature maser\label{Synchro-curvature maser}}
\subsubsection{Radiative transition rate\label{Radiative transition rate}}
Using the wave functions derived in Sec.~\ref{Formulation_Helical magnetic field}, we now calculate the radiative transition rates between different eigenstates.
The interaction Hamiltonian between an electron and a radiation field is given as
\begin{align}
  \hat{\mathcal{H}}_{\mathrm{int}}
  =ec \int \overline{\Psi}_{\text{f}} \gamma^{\mu} \Psi_{\text{i}} \hat{A}_{\mu} d^3 x,
\end{align}
where $\overline{\Psi}=\Psi^\dag\gamma^0$ is the Dirac conjugate; the initial and final states of an electron are denoted by $\Psi_{\text{i}}$ and $\Psi_{\text{f}}$, respectively.
The operator for the radiation field may be expanded as usual:
\begin{align}
  \hat{A}_{\mu}&=\sqrt{\frac{2\pi \hbar}{\varepsilon_0 V}}
  \sum_{\v{k}}\sum_{\sigma}
  \frac{1}{\sqrt{\omega_k}}
  \left(
  a_{\v{k},\epsilon^{(\sigma)}} \epsilon_\mu^{(\sigma)}(\v{k})e^{i\v{k} \cdot \v{x}-\omega_k t}
  \right.
  \nonumber \\
  &\hspace{20mm}+
  \left.
  a_{\v{k},\epsilon^{(\sigma)}}^{\dag} \epsilon_\mu^{*(\sigma)}(\v{k})e^{-i\left(\v{k} \cdot \v{x}-\omega_k t\right)}
  \right)\label{vector_field},
\end{align}
where the four-dimensional polarization vector is $\epsilon_{\mu}^{(\sigma)}=(0,\v{\epsilon}^{(\sigma)})$ in the Coulomb gauge; $a_{\v{k},\epsilon^{(\sigma)}}^{\dag}$ and $a_{\v{k},\epsilon^{(\sigma)}}$ are the creation and annihilation operators of the photon with $\v{k},\epsilon^{(\sigma)}$, respectively; $V$ is an arbitrary volume of quantization and $\varepsilon_0$ is the electric permittivity of vacuum ($\varepsilon_0=1$ in CGS units).

The transition rate is derived from the matrix element squared calculated to the lowest order in the perturbation theory as
\begin{align}
  \frac{2\pi}{\hbar}\left\|\bra{1_{\v{k},\epsilon^{(\sigma)}}} \hat{\mathcal{H}}_{\mathrm{int}}\ket{0}\right\|^2 \delta(E_{n',l'_\parallel}+h \nu-E_{n,l_\parallel}),
\end{align}
where $\ket{0},\ket{1_{\v{k},\epsilon^{(\sigma)}}}$ represent the initial and final photon states, respectively.
We multiply this with the density of state of photons as well as that of electrons.
Then the transition rate from vacuum to a single-photon state per frequency per solid angle per time can be written as follows:
\begin{align}
  \mathcal{W}_{\text{emis}}^{(\sigma)}&=
  \int \frac{L}{2\pi \hbar}dP_{z_{\text{h}}} \frac{V \nu^2}{c^3} \nonumber \\
  &\quad \times \frac{2\pi}{\hbar}\left\|\bra{1_{\v{k},\epsilon^{(\sigma)}}} \hat{\mathcal{H}}_{\mathrm{int}}\ket{0}\right\|^2 \delta(E_{n',l'_\parallel}+h \nu-E_{n,l_\parallel}).
  \label{68}
\end{align}
Setting the quantization length as $L=2\pi R/\cos{\alpha}$, which corresponds to the one round of helix, we get
\begin{align}
  \mathcal{W}_{\text{emis}}^{(\sigma)}&=
  \frac{R}{\hbar c\cos{\alpha}} \frac{V \nu^2}{c^3}
  \frac{2\pi}{\hbar}
  \frac{e^2 \hbar}{\varepsilon_0 V \nu}
  \left\|j^\mu_{\text{emis}} \epsilon_\mu^{(\sigma)}\right\|^2
  \label{69}
  \\
  &=\frac{2\pi Re^2\nu}{\varepsilon_0 \hbar  c^4 \cos{\alpha}}
  \left\|j^\mu_{\text{emis}} \epsilon_\mu^{(\sigma)}\right\|^2,
  \label{emis}\\
  j^\mu_{\text{emis}} &\equiv c\int \overline{\Psi}_{\text{f}} \gamma^\mu \Psi_{\text{i}} e^{-i \v{k}\cdot \v{x}} d^3 x.
\end{align}
In deriving Eq.~\eqref{69} from Eq.~\eqref{68}, we employ the assumption
that the translational motion along the magnetic-field line dominates other contributions in the energy ($E_{n,l_\parallel}\approx c P_{z_{\text{h}}}$).

We can get the absorption rate in the similar way.
In this case, the photon states change as $\ket{1_{\v{k},\epsilon^{(\sigma)}}} \rightarrow \ket{0}$, and the transition rate is written as
\begin{align}
  \mathcal{W}_{\text{abs}}^{(\sigma)}
  &=\frac{2\pi Re^2\nu}{\varepsilon_0 \hbar  c^4 \cos{\alpha}}
  \left\|j^\mu_{\text{abs}} \epsilon_\mu^{(\sigma)}\right\|^2,
  \label{abs}\\
  j^\mu_{\text{abs}} &\equiv c\int \overline{\Psi}_{\text{f}} \gamma^\mu \Psi_{\text{i}} e^{i \v{k}\cdot \v{x}} d^3 x.
\end{align}

\subsubsection{Calculations of transition current\label{Calculations of transition current}}
The wave functions representing the initial and final states of an electron are given in Sec.~\ref{Formulation_Helical magnetic field} as
\begin{subequations}
  \begin{align}
    \Psi_{\text{i}}(\v{r})=
    S^{23}(\theta) S^{31}(\alpha)
    \begin{pmatrix}
      c_1 f_{l_{\perp}-1}(\v{r})\\
      c_2 f_{l_{\perp}}(\v{r})\\
      c_3 f_{l_{\perp}-1}(\v{r})\\
      c_4 f_{l_{\perp}}(\v{r})\\
    \end{pmatrix}
    ,\\
    \Psi_{\text{f}}(\v{r})=
    S^{23}(\theta) S^{31}(\alpha)
    \begin{pmatrix}
      c'_1 f_{l'_{\perp}-1}(\v{r})\\
      c'_2 f_{l'_{\perp}}(\v{r})\\
      c'_3 f_{l'_{\perp}-1}(\v{r})\\
      c'_4 f_{l'_{\perp}}(\v{r})\\
    \end{pmatrix}. \label{initialfinal}
  \end{align}
\end{subequations}
The coefficients $c_1\sim c_4$ and $c'_1\sim c'_4$ depend on the eigenstates we choose.
The above expressions with the spinor rotation matrices are clear and concise, and facilitate the following calculations considerably.

The wave number and polarization vectors are given as
\begin{subequations}
  \begin{align}
    \v{k}&=k\v{k}_0
    =k
    \begin{pmatrix}
      \sin{\kappa}\\
      \cos{\kappa}\cos{\zeta}\\
      \cos{\kappa}\sin{\zeta}\\
    \end{pmatrix}
    ,
    \\
    \v{\epsilon}^{(1)}&=
    \begin{pmatrix}
      0\\
      \sin{\zeta}\\
      -\cos{\zeta}\\
    \end{pmatrix}
    ,\\
    \v{\epsilon}^{(2)}&=\v{k}_0 \times \v{\epsilon}^{(1)}=
    \begin{pmatrix}
      -\cos{\kappa}\\
      \sin{\kappa}\cos{\zeta}\\
      \sin{\kappa}\sin{\zeta}\\
    \end{pmatrix},
  \end{align}
\end{subequations}
where $\kappa$ is the angle measured from the $yz$-plane and $\zeta$ is the angle from the $y$-axis of the projected wave number vector on the $yz$-plane.
In this definition, the polarization vectors are real, $\epsilon_\mu^{(\sigma)}=\epsilon^{*(\sigma)}_\mu$.
In the rest of the paper we consider the case with $s=0$ alone, which would not lose generality \cite{Sokolov1968}.
\begin{widetext}
  \noindent
  % Then the subscripts $l_{\perp}, l'_{\perp}$ in Eq.~\eqref{initialfinal} are changed to $n, n'$ respectively and
  Then the product $j^{\mu}_{\text{emis}}\epsilon_\mu^{(\sigma)}$ can be calculated as follows:
  \begin{align}
    j^{\mu}_{\text{emis}}\epsilon_\mu^{(\sigma)}
    &=
    c
    \int_{-\frac{\pi R}{\cos{\alpha}}}^{\frac{\pi R}{\cos{\alpha}}} d z_{\text{h}}
    \int_0^{\infty} dr \int_{-\pi}^{\pi} d\phi\ r \cdot
    e^{-i\v{k}\cdot \v{x}}
    \begin{pmatrix}
      c_1' f_{n'-1}(\v{r})\\
      c_2' f_{n'}(\v{r})\\
      c_3' f_{n'-1}(\v{r})\\
      c_4' f_{n'}(\v{r}) \\
    \end{pmatrix}
    ^{\dag}
    \gamma^0
    S^{31}(\alpha)^{-1} S^{23}(\theta)^{-1}
    \gamma^\mu
    S^{23}(\theta) S^{31}(\alpha)
    \epsilon_\mu^{(\sigma)}
    \begin{pmatrix}
      c_1 f_{n-1}(\v{r})\\
      c_2 f_{n}(\v{r})\\
      c_3 f_{n-1}(\v{r})\\
      c_4 f_{n}(\v{r})\\
    \end{pmatrix}
    \displaybreak[1] \\
    &=c\frac{\lambda \cos{\alpha}}{2\pi R}
    \left[
    \left(c_1'^{*}c_1-c_3'^{*}c_3\right)I_1
    +\left(c_1'^{*}c_2-c_3'^{*}c_4\right)I_2
    +\left(c_2'^{*}c_1-c_4'^{*}c_3\right)I_3
    +\left(c_2'^{*}c_2-c_4'^{*}c_4\right)I_4
    \right],
    \label{c_I}
  \end{align}
  \begin{subequations}
    \begin{align}
      I_1&\equiv
      \frac{1}{\pi \sqrt{(n-1)!(n'-1)!}}
      \int_{-\frac{\pi R}{\lambda \cos{\alpha}}}^{\frac{\pi R}{\lambda \cos{\alpha}}} du
      \int_0^{\infty} dv \int_{-\pi}^{\pi} d\phi\
      e^{i\Delta l_\parallel u}g_1\left(\theta\right)
      e^{-v^2}v^{n+n'-1}
      e^{i(n-n')\phi}
      \cdot e^{-i\v{k}\cdot \v{x}}, \label{I_1}
      \displaybreak[1]
      \\
      I_2&\equiv
      \frac{1}{\pi \sqrt{n!(n'-1)!}}
      \int_{-\frac{\pi R}{\lambda \cos{\alpha}}}^{\frac{\pi R}{\lambda \cos{\alpha}}} du
      \int_0^{\infty} dv \int_{-\pi}^{\pi} d\phi\
      e^{i\Delta l_\parallel u}g_2\left(\theta\right)
      e^{-v^2}v^{n+n'}
      e^{i(n-n'+1)\phi}
      \cdot e^{-i\v{k}\cdot \v{x}},
      \displaybreak[1]
      \\
      I_3&\equiv
      \frac{1}{\pi \sqrt{(n-1)!n'!}}
      \int_{-\frac{\pi R}{\lambda \cos{\alpha}}}^{\frac{\pi R}{\lambda \cos{\alpha}}} du
      \int_0^{\infty} dv \int_{-\pi}^{\pi} d\phi\
      e^{i\Delta l_\parallel u}g_3\left(\theta\right)
      e^{-v^2}v^{n+n'}
      e^{i(n-n'-1)\phi}
      \cdot e^{-i\v{k}\cdot \v{x}},
      \displaybreak[1]
      \\
      I_4&\equiv
      \frac{1}{\pi \sqrt{n!n'!}}
      \int_{-\frac{\pi R}{\lambda \cos{\alpha}}}^{\frac{\pi R}{\lambda \cos{\alpha}}} du
      \int_0^{\infty} dv \int_{-\pi}^{\pi} d\phi\
      e^{i\Delta l_\parallel u}g_4\left(\theta\right)
      e^{-v^2}v^{n+n'+1}
      e^{i(n-n')\phi}
      \cdot e^{-i\v{k}\cdot \v{x}}. \label{I_4}
    \end{align}
  \end{subequations}
\end{widetext}
In so doing, we employ the following equation:
\begin{gather}
  \gamma^0 S^{31}(\alpha)^{-1}  S^{23}(\theta)^{-1} \gamma^\mu S^{23}(\theta) S^{31}(\alpha) \epsilon_{\mu}^{(\sigma)}
  \nonumber\\
  \equiv
  \begin{pmatrix}
    g_1 (\theta)  & g_2 (\theta)  & 0 & 0\\
    g_3 (\theta)  & g_4 (\theta)  & 0 & 0\\
    0 & 0 & -g_1 (\theta)  & -g_2 (\theta) \\
    0 & 0 & -g_3 (\theta)  & -g_4 (\theta) \\
  \end{pmatrix},\label{eq1}
\end{gather}
where the functions $g_1(\theta)\sim g_4(\theta)$ depend on the polarization vector.
The integration range for $u$ corresponds to one round of helix.

We first evaluate $e^{i\Delta l_\parallel u}e^{-i\v{k}\cdot \v{x}}$ following \cite{Voisin2017}.
For highly relativistic electrons with the momentum along the magnetic-field line being dominant, the energy can be expanded as follows:
  \begin{align}
    E_{n,l_\parallel}
    &\approx \frac{\hbar c l_\parallel}{\lambda}
    \left(
    1+\frac{1}{2\gamma^2}+\frac{1}{\gamma^2}\frac{B}{B_{\text{c}}}n
    +\mathcal{O} \left(\frac{1}{\gamma^4}\right)
    \right),\label{energy_approx}
  \end{align}
where $\gamma \equiv E_{n,l_\parallel}/mc^2$ is the Lorentz factor.
This is applied to both the initial and final states as
\begin{subequations}
  \begin{align}
    E_{n,l_\parallel}&
    \approx
    \frac{\hbar c l_\parallel}{\lambda}\left[1+\frac{1}{2\gamma^2}\left(1+2\frac{B}{B_{\text{c}}}n\right)\right],\\
    E_{n',l'_\parallel}&
    \approx
    \frac{\hbar c l'_\parallel}{\lambda}\left[1+\frac{1}{2\gamma '^2}\left(1+2\frac{B}{B_{\text{c}}}n'\right)\right],
  \end{align}
\end{subequations}
where the unprimed and primed variables are concerning the initial and final states, respectively.
The energy conservation $E_{n,l_\parallel}-E_{n',l'_\parallel}=h \nu$ leads to the following relations:
\begin{gather}
  \Delta l_\parallel=\lambda k
  \left(
  1+\frac{1}{2\gamma_2}
  \right),
  \\
  \frac{1}{\gamma_2}
  \equiv
  \frac{1}{\gamma^2}\left[
  \left(1+\frac{h\nu}{E_{n,l_\parallel}}\right)
  \left(1+2\frac{B}{B_{\text{c}}}n'\right)
  -
  2\frac{E_{n,l_\parallel}}{h \nu}\frac{B}{B_{\text{c}}}\Delta n
  \right],
\end{gather}
where $\Delta l_{\parallel} =l_{\parallel}-l'_{\parallel}$ and $\Delta n =n-n'$ (see Appendix \ref{C} for detailed calculations).
The inner product $\v{k}\cdot \v{x}$ is calculated as
\begin{align}
  \v{k}\cdot \v{x}
  &=k\left(z_{\text{h}}\sin{\alpha}+r\cos{\phi}\cos{\alpha}\right)
  \sin{\kappa}
  \nonumber\\
  &\quad +k\left(R+r\sin{\phi}\right)\cos{\kappa} \cos{\left(\zeta-\theta\right)}
  \nonumber\\
  &\quad -kr\cos{\phi}\sin{\alpha} \cos{\kappa} \sin{\left(\zeta-\theta\right)}.
  \label{kx}
\end{align}
Hereafter we set $\zeta=\pi/2$ without loss of generality.
Then we obtain the followings:
% \begin{subequations}
  \begin{align}
    \v{k}
    =k
    \begin{pmatrix}
      \sin{\kappa}\\
      0\\
      \cos{\kappa}\\
    \end{pmatrix}
    ,
    % \\
    \v{\epsilon}^{(1)}
    =
    \begin{pmatrix}
      0\\
      1\\
      0\\
    \end{pmatrix}
    ,
    % \\
    \v{\epsilon}^{(2)}
    =
    \begin{pmatrix}
      -\cos{\kappa}\\
      0\\
      \sin{\kappa}\\
    \end{pmatrix},
  \end{align}
% \end{subequations}
\begin{subequations}
  \begin{align}
    \begin{cases}
      g_1 (\theta) =-\sin{\theta}\\
      g_2 (\theta) =-i\cos{\theta}\\
      g_3 (\theta) =i\cos{\theta}\\
      g_4 (\theta) =\sin{\theta}\\
    \end{cases}  (\text{for }\v{\epsilon}^{(1)}), \label{g_1}
  \end{align}
and
  \begin{align}
    \begin{cases}
      g_1 (\theta) =
      -\cos{\kappa}\sin{\alpha}\cos{\theta}+\sin{\kappa}\cos{\alpha}\cos{\theta}
      \\
      g_2 (\theta) =
      -\cos{(\kappa-\alpha)}-i\sin{(\kappa-\alpha)}\sin{\theta}
      \\
      g_3 (\theta) =
      -\cos{(\kappa-\alpha)}+i\sin{(\kappa-\alpha)}\sin{\theta}
      \\
      g_4 (\theta) =
      \cos{\kappa}\sin{\alpha}\cos{\theta}-\sin{\kappa}\cos{\alpha}\cos{\theta}
      \\
    \end{cases}  (\text{for }\v{\epsilon}^{(2)}). \label{g_2}
  \end{align}
\end{subequations}
Using the nondimensionalized variables, Eq.~\eqref{kx} is rewritten as
\begin{align}
  \v{k}\cdot \v{x}
  &=\lambda k
    \left(u\sin{\alpha}+v\cos{\phi}\cos{\alpha}\right)
    \sin{\kappa}
  \nonumber \\
  &\quad +\rho k \cos^2{\alpha} \cos{\kappa} \sin{\theta} +\lambda kv\sin{\phi} \cos{\kappa} \sin{\theta}
  \nonumber \\
  &\quad -\lambda kv \cos{\phi}\sin{\alpha} \cos{\kappa} \cos{\theta}.
\end{align}
Then we obtain
\begin{align}
  &e^{i\Delta l_\parallel u}e^{-i\v{k}\cdot \v{x}} \nonumber\\
  &=e^{iH(\theta)}e^{-i \lambda kv
  \left[\left( \cos{\alpha} \sin{\kappa}-\sin{\alpha} \cos{\kappa}
  \cos{\theta} \right) \cos{\phi}
  + \cos{\kappa}\sin{\theta} \sin{\phi}\right]},
  \label{taylor_before}
  \displaybreak[1]
  \\
  &H(\theta)\equiv
  \rho k
  \left[
  \left(1+\frac{1}{2\gamma_2}-\sin{\alpha}\sin{\kappa}\right)\theta\cos{\alpha}
  \right. \nonumber \\
  &\hspace{40mm}-\cos^2{\alpha}\cos{\kappa}\sin{\theta}
  \biggr].  \label{taylor}
\end{align}
We note that the variable change $u\rightarrow \theta$ leads to
\begin{align}
  \int_{-\frac{\pi R}{\lambda \cos{\alpha}}}^{\frac{\pi R}{\lambda \cos{\alpha}}} du
  =\frac{R}{\lambda \cos{\alpha}}\int_{-\pi}^{\pi} d\theta.
\end{align}

Now we perform the integrations in turn.
We start with the $\phi$-integration.
In so doing, we expand Eq.~\eqref{taylor} in $v$ as $\lambda k$ is expected to be small:
\begin{align}
  \lambda k = \sqrt{\frac{2\hbar c}{eB}} \frac{2\pi \nu}{c} =\sqrt{\frac{4\pi h}{ec}}\frac{\nu}{\sqrt{B}}
  \sim 10^{-6} \times\frac{\nu_9}{\sqrt{B_2}},\label{taylor_order}
\end{align}
where we have in mind the application to the GHz maser in the pulsar magnetosphere and $\nu_{9}$ and $B_2$ are defined as $\nu=\nu_{9}\times10^{9}\text{Hz}$ and $B=B_2\times10^{2}\text{G}$, respectively.
\begin{widetext}
  Then Eq.~\eqref{c_I} is rewritten as follows:
  \begin{gather}
    j^{\mu}_{\text{emis}}\epsilon_\mu^{(\sigma)}
    =\frac{c}{2\pi}
    \left[
    \left(c_1'^{*}c_1-c_3'^{*}c_3\right)I'_1
    +\left(c_1'^{*}c_2-c_3'^{*}c_4\right)I'_2
    +\left(c_2'^{*}c_1-c_4'^{*}c_3\right)I'_3
    +\left(c_2'^{*}c_2-c_4'^{*}c_4\right)I'_4
    \right],
    \label{j_mu}
  \end{gather}
  \begin{subequations}
    \begin{align}
      I'_1&=
      \frac{1}{\pi\sqrt{(n-1)!(n'-1)!}}\int_{-\pi}^{\pi} d\theta\
      e^{iH(\theta)}
      g_1(\theta)\nonumber \\
      &
      \int_0^{\infty} dv \int_{-\pi}^\pi d\phi\
      e^{-v^2}v^{n+n'-1}
      e^{i(n-n')\phi}
      \sum_{q=0}^{\infty}\frac{(-i \lambda kv)^q}{q!}
      \left[\left( \cos{\alpha} \sin{\kappa}-\sin{\alpha} \cos{\kappa} \cos{\theta} \right) \cos{\phi}
      + \cos{\kappa}\sin{\theta} \sin{\phi}\right]^q,
      \label{I'_1}
      \displaybreak[1] \\
      I'_2&=
      \frac{1}{\pi\sqrt{n!(n'-1)!}}\int_{-\pi}^{\pi} d\theta\
      e^{iH(\theta)}
      g_2(\theta)\nonumber \\
      &
      \int_0^{\infty} dv \int_{-\pi}^\pi d\phi\
      e^{-v^2}v^{n+n'}
      e^{i(n-n'+1)\phi}
      \sum_{q=0}^{\infty}\frac{(-i \lambda kv)^q}{q!}
      \left[\left( \cos{\alpha} \sin{\kappa}-\sin{\alpha} \cos{\kappa} \cos{\theta} \right) \cos{\phi}
      + \cos{\kappa}\sin{\theta} \sin{\phi}\right]^q,
      \label{I'_2}
      \displaybreak[1] \\
      I'_3&=
      \frac{1}{\pi\sqrt{(n-1)!n'!}}\int_{-\pi}^{\pi} d\theta\
      e^{iH(\theta)}
      g_3(\theta)\nonumber \\
      &
      \int_0^{\infty} dv \int_{-\pi}^\pi d\phi\
      e^{-v^2}v^{n+n'}
      e^{i(n-n'-1)\phi}
      \sum_{q=0}^{\infty}\frac{(-i \lambda kv)^q}{q!}
      \left[\left( \cos{\alpha} \sin{\kappa}-\sin{\alpha} \cos{\kappa} \cos{\theta} \right) \cos{\phi}
      + \cos{\kappa}\sin{\theta} \sin{\phi}\right]^q,
      \label{I'_3}
      \displaybreak[1] \\
      I'_4&=
      \frac{1}{\pi\sqrt{n!n'!}}\int_{-\pi}^{\pi} d\theta\
      e^{iH(\theta)}
      g_4(\theta)\nonumber \\
      &
      \int_0^{\infty} dv \int_{-\pi}^\pi d\phi\
      e^{-v^2}v^{n+n'+1}
      e^{i(n-n')\phi}
      \sum_{q=0}^{\infty}\frac{(-i \lambda kv)^q}{q!}
      \left[\left( \cos{\alpha} \sin{\kappa}-\sin{\alpha} \cos{\kappa} \cos{\theta} \right) \cos{\phi}
      + \cos{\kappa}\sin{\theta} \sin{\phi}\right]^q.
      \label{I'_4}
    \end{align}
  \end{subequations}
\end{widetext}

Then the integration with respect to $\phi$ is reduced to the following form:
\begin{align}
  & A_{pq}
   \equiv \int_{-\pi}^{\pi} d\phi e^{ip\phi} \left(a\cos{\phi}+b\sin{\phi}\right)^q,
  \\
  a&=-i \lambda kv\left( \cos{\alpha} \sin{\kappa}-\sin{\alpha} \cos{\kappa}
  \cos{\theta} \right), \nonumber
  \\
  b&=-i \lambda kv \cos{\kappa} \sin{\theta}. \nonumber
\end{align}
We can calculate $A_{pq}$ analytically.
In fact it does not vanish only in some limited combinations of $p$ and $q$: for a fixed $p$, $q$ should satisfy $q \geq |p|$.
Note that the value of $q$ is the power of $\lambda k$.
Considering up to the second order, we obtain non-vanishing $A_{pq}$ as follows:
\begin{align}
  A_{pq}
  =
  \begin{dcases}
    2\pi  &(p=0,q=0)\\
    \pi(a^2+b^2)  & (p=0,q=2)\\
    \pi(a\pm ib)  & (p=\pm1,q=1)\\
    \frac{\pi}{2}(a\pm ib)^2  & (p=\pm2,q=2)
  \end{dcases}.\label{A_pq}
\end{align}
Then the $v$-integration can be accomplished just by using the following formula:
\begin{align}
  \int_0^{\infty} e^{-v^2}v^{2m+1}dv=\frac{m!}{2} \quad (m\in{\mathbb{N}}).
\end{align}
The remaining $\theta$-intagrals in Eqs.~\eqref{I'_1}$\sim$\eqref{I'_4} are now reduced to the following form:
\begin{align}
  B_{pq}\equiv \int_{-\pi}^{\pi} e^{i H(\theta)} (\cos{\theta})^p (\sin{\theta})^q  d\theta,
\end{align}
which cannot be done analytically and will be evaluated numerically later.
We have so far considered photon emissions.
For absorptions, the energy conservation is given as $E_{n,l_\parallel}-E_{n',l'_\parallel}=-h \nu$, implying that we have only to make the following substitutions: $\nu$ for $-\nu$ and $k$ for $-k$.\\

Below we give the concrete expressions of $I'_1\sim I'_4$ in $j^{\mu}_{\text{emis}}\epsilon_\mu^{(\sigma)}$ for each polarization and the following transitions between the Landau levels: $n\rightarrow n$, $n\rightarrow n-1$, $n\rightarrow n+1$, which are consistent with the expansion in $\lambda k$ up to the second order.
\begin{widetext}
	\begin{align*}
	\bullet & \  n\rightarrow n,\v{\epsilon}^{(1)} \nonumber\\
		&I'_1=
		-B_{01}+\frac{n\lambda^2 k^2}{4}
		\left(\cos^2{\alpha}\sin^2{\kappa} B_{01}
		-2\sin{\alpha}\cos{\alpha}\sin{\kappa}\cos{\kappa}B_{11}
		+\sin^2{\alpha} \cos^2{\kappa}B_{21}
		+\cos^2{\kappa}B_{03}\right)
		\displaybreak[1] \\
		&I'_2=
		\frac{\sqrt{n}\lambda k}{2}
		\left(-\cos{\alpha}\sin{\kappa}B_{10}
		+\sin{\alpha}\cos{\kappa}B_{20}
		-i\cos{\kappa}B_{11}\right)
		\displaybreak[1] \\
		&I'_3=
		\frac{\sqrt{n}\lambda k}{2}
		\left(\cos{\alpha}\sin{\kappa}B_{10}
		-\sin{\alpha}\cos{\kappa}B_{20}
		-i\cos{\kappa}B_{11}\right)
		\displaybreak[1] \\
		&I'_4=
		B_{01}
		-\frac{(n+1)\lambda^2 k^2}{4}
		\left(\cos^2{\alpha}\sin^2{\kappa} B_{01}
		-2\sin{\alpha}\cos{\alpha}\sin{\kappa}\cos{\kappa}B_{11}
		+\sin^2{\alpha} \cos^2{\kappa}B_{21}
		+\cos^2{\kappa}B_{03}\right)
		\\
		\displaybreak[1] \\
	\bullet & \  n\rightarrow n-1,\v{\epsilon}^{(1)} \nonumber\\
		&I'_1=
		\frac{\sqrt{n-1}\lambda k}{2}
		\left(
		i\cos{\alpha}\sin{\kappa}B_{01}
		-i\sin{\alpha}\cos{\kappa}B_{11}
		-\cos{\kappa}B_{02}
		\right)
		\displaybreak[1] \\
		&I'_2=
		\frac{\sqrt{n(n-1)}\lambda^2 k^2}{4}
		\left(
		i\cos^2{\alpha}\sin^2{\kappa}B_{10}
		-2i\sin{\alpha}\cos{\alpha}\sin{\kappa}\cos{\kappa}B_{20}
		+i\sin^2{\alpha}\cos^2{\kappa}B_{30}
		\right.\nonumber
		\\
		&\hspace{80mm}\left.
		-2\cos{\alpha}\sin{\kappa}\cos{\kappa}B_{11}
		+2\sin{\alpha}\cos^2{\kappa}B_{21}
		-i\cos^2{\kappa}B_{12}
		\right)
		\displaybreak[1] \\
		&I'_3=
		iB_{10}-\frac{n\lambda^2 k^2}{4}
		\left(i\cos^2{\alpha}\sin^2{\kappa}B_{10}
		-2i\sin{\alpha}\cos{\alpha}\sin{\kappa}\cos{\kappa}B_{20}
		+i\sin^2{\alpha}\cos^2{\kappa}B_{30}
		+i\cos^2{\kappa}B_{12}\right)
		\displaybreak[1] \\
		&I'_4=
		\frac{\sqrt{n}\lambda k}{2}
		\left(
		-i\cos{\alpha}\sin{\kappa}B_{01}
		+i\sin{\alpha}\cos{\kappa}B_{11}
		+\cos{\kappa}B_{02}
		\right)
		\\
		\displaybreak[1] \\
	\bullet & \  n\rightarrow n+1,\v{\epsilon}^{(1)} \nonumber\\
		&I'_1=
		\frac{\sqrt{n}\lambda k}{2}
		\left(
		i\cos{\alpha}\sin{\kappa}B_{01}
		-i\sin{\alpha}\cos{\kappa}B_{11}
		+\cos{\kappa}B_{02}
		\right)
		\displaybreak[1] \\
		&I'_2=
		-iB_{10}+\frac{(n+1)\lambda^2 k^2}{4}
		\left(i\cos^2{\alpha}\sin^2{\kappa}B_{10}
		-2i\sin{\alpha}\cos{\alpha}\sin{\kappa}\cos{\kappa}B_{20}
		+i\sin^2{\alpha}\cos^2{\kappa}B_{30}
		+i\cos^2{\kappa}B_{12}\right)
		\displaybreak[1] \\
		&I'_3=
		\frac{\sqrt{n(n+1)}\lambda^2 k^2}{4}
		\left(
		-i\cos^2{\alpha}\sin^2{\kappa}B_{10}
		+2i\sin{\alpha}\cos{\alpha}\sin{\kappa}\cos{\kappa}B_{20}
		-i\sin^2{\alpha}\cos^2{\kappa}B_{30}
		\right.\nonumber
		\\
		&\hspace{80mm}\left.
		-2\cos{\alpha}\sin{\kappa}\cos{\kappa}B_{11}
		+2\sin{\alpha}\cos^2{\kappa}B_{21}
		+i\cos^2{\kappa}B_{12}
		\right)
		\displaybreak[1] \\
		&I'_4=
		\frac{\sqrt{n+1}\lambda k}{2}
		\left(
		-i\cos{\alpha}\sin{\kappa}B_{01}
		+i\sin{\alpha}\cos{\kappa}B_{11}
		-\cos{\kappa}B_{02}
		\right)
		\\
		\displaybreak[1] \\
	\bullet & \  n\rightarrow n,\v{\epsilon}^{(2)}\\
		&I'_1=
		\sin{(\kappa-\alpha)}
		\left[B_{10}-\frac{n\lambda^2 k^2}{4}
		\left(\cos^2{\alpha}\sin^2{\kappa} B_{10}
		-2\sin{\alpha}\cos{\alpha}\sin{\kappa}\cos{\kappa}B_{20}
		+\sin^2{\alpha} \cos^2{\kappa}B_{30}
		+\cos^2{\kappa}B_{12}\right)\right]
		\displaybreak[1] \\
		&I'_2=
		\frac{\sqrt{n}\lambda k}{2}
		\cos{(\kappa-\alpha)}
		\left(
		i\cos{\alpha}\sin{\kappa}B_{00}
		-i\sin{\alpha}\cos{\kappa}B_{10}
		-\cos{\kappa}B_{01}
		\right)
		\nonumber
		\\
		&\hspace{70mm}
		+
		\frac{\sqrt{n}\lambda k}{2}
		\sin{(\kappa-\alpha)}
		\left(
		-\cos{\alpha}\sin{\kappa}B_{01}
		+\sin{\alpha}\cos{\kappa}B_{11}
		-i\cos{\kappa}B_{02}
		\right)
		\displaybreak[1] \\
		&I'_3=
		\frac{\sqrt{n}\lambda k}{2}
		\cos{(\kappa-\alpha)}
		\left(
		i\cos{\alpha}\sin{\kappa}B_{00}
		-i\sin{\alpha}\cos{\kappa}B_{10}
		+\cos{\kappa}B_{01}
		\right)
		\nonumber \\
		&\hspace{70mm}
		+
		\frac{\sqrt{n}\lambda k}{2}
		\sin{(\kappa-\alpha)}
		\left(
		\cos{\alpha}\sin{\kappa}B_{01}
		-\sin{\alpha}\cos{\kappa}B_{11}
		-i\cos{\kappa}B_{02}
		\right)
		\displaybreak[1] \\
		&I'_4=
		\sin{(\kappa-\alpha)}
		\left[-B_{10}
		+\frac{(n+1)\lambda^2 k^2}{4}
		\left(\cos^2{\alpha}\sin^2{\kappa} B_{10}
		-2\sin{\alpha}\cos{\alpha}\sin{\kappa}\cos{\kappa}B_{20}
		+\sin^2{\alpha} \cos^2{\kappa}B_{30}
		+\cos^2{\kappa}B_{12}\right)\right]
		\\
		\displaybreak[1] \\
	\bullet & \  n\rightarrow n-1,\v{\epsilon}^{(2)} \nonumber\\
		& I'_1=
		\frac{\sqrt{n-1}\lambda k}{2}
		\sin{(\kappa-\alpha)}
		\left(
		-i\cos{\alpha}\sin{\kappa}B_{10}
		+i\sin{\alpha}\cos{\kappa}B_{20}
		+\cos{\kappa}B_{11}
		\right)
		\displaybreak[1] \\
		& I'_2=
		\frac{\sqrt{n(n-1)}\lambda^2 k^2}{4}
		\cos{(\kappa-\alpha)}
		\left(
		\cos^2{\alpha}\sin^2{\kappa}B_{00}
		-2\sin{\alpha}\cos{\alpha}\sin{\kappa}\cos{\kappa}B_{10}
		+\sin^2{\alpha}\cos^2{\kappa}B_{20}
		\right.
		\nonumber\\
		&\hspace{90mm}\left.
		+2i\cos{\alpha}\sin{\kappa}\cos{\kappa}B_{01}
		-2i\sin{\alpha}\cos^2{\kappa}B_{11}
		-\cos^2{\kappa}B_{02}
		\right)
		\nonumber\\
		&\quad +
		\frac{\sqrt{n(n-1)}\lambda^2 k^2}{4}
		\sin{(\kappa-\alpha)}
		\left(
		i\cos^2{\alpha}\sin^2{\kappa}B_{01}
		-2i\sin{\alpha}\cos{\alpha}\sin{\kappa}\cos{\kappa}B_{11}
		+i\sin^2{\alpha}\cos^2{\kappa}B_{21}
		\right.
		\nonumber \\
		&\hspace{90mm}\left.
		-2\cos{\alpha}\sin{\kappa}\cos{\kappa}B_{02}
		+2\sin{\alpha}\cos^2{\kappa}B_{12}
		-i\cos^2{\kappa}B_{03}
		\right)
		\displaybreak[1] \\
		& I'_3=
		\cos{(\kappa-\alpha)}
		\left[
		-B_{00}+\frac{n\lambda^2 k^2}{4}
		\left(\cos^2{\alpha}\sin^2{\kappa}B_{00}
		-2\sin{\alpha}\cos{\alpha}\sin{\kappa}\cos{\kappa}B_{10}
		+\sin^2{\alpha}\cos^2{\kappa}B_{20}
		+\cos^2{\kappa}B_{02}\right)
		\right]
		\nonumber \\
		&\quad +
		\sin{(\kappa-\alpha)}
		\left[
		iB_{01}-\frac{n\lambda^2 k^2}{4}
		\left(i\cos^2{\alpha}\sin^2{\kappa}B_{01}
		-2i\sin{\alpha}\cos{\alpha}\sin{\kappa}\cos{\kappa}B_{11}
		+i\sin^2{\alpha}\cos^2{\kappa}B_{21}
		+i\cos^2{\kappa}B_{03}\right)
		\right]
		\displaybreak[1] \\
		& I'_4=
		\frac{\sqrt{n}\lambda k}{2}
    \sin{(\kappa-\alpha)}
		\left(
		i\cos{\alpha}\sin{\kappa}B_{10}
		-i\sin{\alpha}\cos{\kappa}B_{20}
		-\cos{\kappa}B_{11}
		\right)
		\\
		\displaybreak[1] \\
	\bullet & \  n\rightarrow n+1,\v{\epsilon}^{(2)} \nonumber\\
		& I'_1=
		\frac{\sqrt{n}\lambda k}{2}
		\sin{(\kappa-\alpha)}
		\left(
		-i\cos{\alpha}\sin{\kappa}B_{10}
		+i\sin{\alpha}\cos{\kappa}B_{20}
		-\cos{\kappa}B_{11}
		\right)
		\displaybreak[1] \\
		& I'_2=
		\cos{(\kappa-\alpha)}
		\left[
		-B_{00}+\frac{(n+1)\lambda^2 k^2}{4}
		\left(\cos^2{\alpha}\sin^2{\kappa}B_{00}
		-2\sin{\alpha}\cos{\alpha}\sin{\kappa}\cos{\kappa}B_{10}
		+\sin^2{\alpha}\cos^2{\kappa}B_{20}
		+\cos^2{\kappa}B_{02}\right)
		\right]
		\nonumber \\
		&\quad+
		\sin{(\kappa-\alpha)}
		\left[
		-iB_{01}+\frac{(n+1)\lambda^2 k^2}{4}
		\left(i\cos^2{\alpha}\sin^2{\kappa}B_{01}
		-2i\sin{\alpha}\cos{\alpha}\sin{\kappa}\cos{\kappa}B_{11}
		\right.
		\right.
		\nonumber
		\\
		&\hspace{122mm}
		\left.
		\left.
		+i\sin^2{\alpha}\cos^2{\kappa}B_{21}
		+i\cos^2{\kappa}B_{03}\right)\right]
		\displaybreak[1] \\
		& I'_3=
		\frac{\sqrt{n(n+1)}\lambda^2 k^2}{4}
		\cos{(\kappa-\alpha)}
		\left(
		\cos^2{\alpha}\sin^2{\kappa}B_{00}
		-2\sin{\alpha}\cos{\alpha}\sin{\kappa}\cos{\kappa}B_{10}
		+\sin^2{\alpha}\cos^2{\kappa}B_{20}
		\right.
		\nonumber \\
		&\hspace{90mm}\left.
		-2i\cos{\alpha}\sin{\kappa}\cos{\kappa}B_{01}
		+2i\sin{\alpha}\cos^2{\kappa}B_{11}
		-\cos^2{\kappa}B_{02}
		\right)
		\nonumber \\
		&\quad+
		\frac{\sqrt{n(n+1)}\lambda^2 k^2}{4}
		\sin{(\kappa-\alpha)}
		\left(
		-i\cos^2{\alpha}\sin^2{\kappa}B_{01}
		+2i\sin{\alpha}\cos{\alpha}\sin{\kappa}\cos{\kappa}B_{11}
		-i\sin^2{\alpha}\cos^2{\kappa}B_{21}
		\right.
		\nonumber \\
		&\hspace{90mm}\left.
		-2\cos{\alpha}\sin{\kappa}\cos{\kappa}B_{02}
		+2\sin{\alpha}\cos^2{\kappa}B_{12}
		+i\cos^2{\kappa}B_{03}
		\right)
		\displaybreak[1] \\
		& I'_4=
		\frac{\sqrt{n+1}\lambda k}{2}
    \sin{(\kappa-\alpha)}
		\left(
		i\cos{\alpha}\sin{\kappa}B_{10}
		-i\sin{\alpha}\cos{\kappa}B_{20}
		+\cos{\kappa}B_{11}
		\right)
	\end{align*}

\subsubsection{Coefficients of eigenstates\label{Coefficients of eigenstates}}
	In the next section, where we evaluate the radiative transition rates numerically, we adopt the eigenstates of the spin operator along the magnetic field (see Sec.~\ref{Helicity and spin eigenstates}).
	Since we assume in this paper that the electron moves highly relativistically along the magnetic field, the coefficients of the eigenstates can be calculated as follows.
	First, $K_1$ and $K_2$ in Eqs.~\eqref{K1} and \eqref{K2} are approximated as
	\begin{subequations}
		\begin{align}
			K_1 &=\frac{\sqrt{m^2c^4+\left(cP_z\right)^2}}{E}
			\approx
			1-\frac{1}{\gamma^2}\frac{B}{B_c}n
			,\\
			K_2 &=\frac{cP_z}{\sqrt{m^2c^4+\left(cP_z\right)^2}}
			\approx
			1-\frac{1}{2\gamma^2}.
		\end{align}
	\end{subequations}
	The coefficients $c_1\sim c_4$ are given, on the other hand, as
	\begin{subequations}
		\begin{align}
			&
			\begin{dcases}
				c_{1 \uparrow}\approx
				\frac{1}{2}
				\sqrt{2}
				\sqrt{2}
				=1,
				\\
				c_{2 \uparrow}\approx
				\frac{i}{2}
				\sqrt{\frac{1}{\gamma^2}\frac{B}{B_c}n}
				\sqrt{2}
				=\frac{i}{\gamma}\sqrt{\frac{B}{2 B_c}n},
				\\
				c_{3 \uparrow}\approx
				\frac{1}{2}
				\sqrt{2}
				\sqrt{\frac{1}{2\gamma^2}}
				=\frac{1}{2\gamma},
				\\
				c_{4 \uparrow}\approx
				-\frac{i}{2}
				\sqrt{\frac{1}{\gamma^2}\frac{B}{B_c}n}
				\sqrt{\frac{1}{2\gamma^2}}
				=-\frac{i}{2\gamma^2}\sqrt{\frac{B}{2B_c}n},
			\end{dcases}
			,
			\displaybreak[1] \\
			&
			\begin{dcases}
				c_{1 \downarrow}\approx
				\frac{1}{2}
				\sqrt{\frac{1}{\gamma^2}\frac{B}{B_c}n}
				\sqrt{\frac{1}{2\gamma^2}}
				=\frac{1}{2\gamma^2}\sqrt{\frac{B}{2B_c}n},
				\\
				c_{2 \downarrow}\approx
				\frac{i}{2}
				\sqrt{2}
				\sqrt{\frac{1}{2\gamma^2}}
				=\frac{i}{2\gamma},
				\\
				c_{3 \downarrow}\approx
				-\frac{1}{2}
				\sqrt{\frac{1}{\gamma^2}\frac{B}{B_c}n}
				\sqrt{2}
				=-\frac{1}{\gamma}\sqrt{\frac{B}{2 B_c}n},
				\\
				c_{4 \downarrow}\approx
				\frac{i}{2}
				\sqrt{2}
				\sqrt{2}
				=i,
			\end{dcases},
		\end{align}
	\end{subequations}
	where the subscripts $\uparrow$ and $\downarrow$ represent the parallel and antiparallel spins along the magnetic field, respectively.
	There are hence four types of spin transition: from $\uparrow$ to $\uparrow$, from $\downarrow$ to $\downarrow$, from $\uparrow$ to $\downarrow$, from $\downarrow$ to $\uparrow$, and the coefficients of relevance in the transition rates are given as follows:
	\begin{gather}
		\begin{array}{c|c|c|c|c}
			& \text{from} \uparrow \text{to} \uparrow
      & \text{from} \downarrow \text{to} \downarrow
      & \text{from} \uparrow \text{to} \downarrow
      & \text{from} \downarrow \text{to} \uparrow
      \rule[-2mm]{0mm}{3mm}  \\ \hline
			\quad c_1'^{*}c_1-c_3'^{*}c_3\quad  &
			1 & \quad -\dfrac{1}{2\gamma^2}\dfrac{B}{B_c}\sqrt{n'n}\quad & \quad \dfrac{1}{\gamma^2}\sqrt{\dfrac{B}{2 B_c}n'}\quad & \quad \dfrac{1}{\gamma^2}\sqrt{\dfrac{B}{2 B_c}n}\quad
			\rule[-4mm]{0mm}{10mm} \\ \hline
			c_1'^{*}c_2-c_3'^{*}c_4 &
			\dfrac{i}{\gamma}\sqrt{\dfrac{B}{2B_c}n} & \dfrac{i}{\gamma}\sqrt{\dfrac{B}{2B_c}n'} & 0 & 0
			\rule[-4mm]{0mm}{10mm} \\ \hline
			c_2'^{*}c_1-c_4'^{*}c_3 &
			-\dfrac{i}{\gamma}\sqrt{\dfrac{B}{2B_c}n'} & -\dfrac{i}{\gamma}\sqrt{\dfrac{B}{2B_c}n} & 0 & 0
			\rule[-4mm]{0mm}{10mm} \\ \hline
			c_2'^{*}c_2-c_4'^{*}c_4 &
			\quad\dfrac{1}{2\gamma^2}\dfrac{B}{B_c}\sqrt{n'n}\quad & -1 & \dfrac{1}{\gamma^2}\sqrt{\dfrac{B}{2 B_c}n} & \dfrac{1}{\gamma^2}\sqrt{\dfrac{B}{2 B_c}n'}
			\rule[-4mm]{0mm}{10mm} \\
		\end{array}\label{Coefficients of spin transitions}
	\end{gather}
Here we neglect the terms that are sub-leading in $\gamma^2$.
It is found that the currents of spin flip transitions are proportional to $1/\gamma^2$ and hence are indeed sub-dominant.
\end{widetext}

\section{Results and Discussion\label{Results and Discussion}}
Based on our formulation given above, we evaluate the true absorption rate numerically.
We are particularly interested in the possibility of maser, which is indicated by the occurrence of a negative true absorption rate.
We have many parameters to set: the environmental variables such as the field strength, curvature radius and pitch angle as well as the variables that characterize electron's motion such as the Lorentz factor and Landau levels.
In this paper we set the values of these parameters to those characteristic of the magnetosphere of NS.
We focus, in particular, on the outer gap of NS.
Then the light cylinder radius $R_{\text{LC}}$, which is written in terms of the rotation period $P$ as
\begin{align}
  R_{\text{LC}}=\frac{cP}{2\pi},
\end{align}
roughly gives the curvature scale of field lines.
In the dipolar magnetic field, which we assume here, the field strength $B$ decreases with the distance $R$ from NS as
\begin{align}
  B=B_{*}\left(\frac{R}{R_{*}}\right)^{-3},
\end{align}
where $B_{*}$ is magnetic-field strength on the NS surface and $R_{*}$ is the radius of NS, which is set to $10^6$cm.
The reference number density in the NS magnetosphere is obtained from the so-called Goldreich-Julian density \cite{Goldreich1969} as
\begin{align}
  n_{\text{e}}=7.00\times 10^{-2} \times \frac{B (\text{G})}{P (\text{s})} \text{cm}^{-3}.
\end{align}
The actual number density of electrons is given by multiplying this number with the multiplicity, which accounts for pair productions.
According to \cite{Medin2010, Harding2011}, the multiplicity is $\sim 10^2-10^4$ in our situation.

We summarize the typical values of the parameters in Table~\ref{Pulsar}.
In addition to the ordinary radio pulsars, we consider millisecond pulsars and magnetars.
For the latter, in particular, we pick up one of the most well-observed magnetars, SGR 1935+2154, and take its observed values \cite{Israel2016}.

\begin{table*}
  \caption{\label{Pulsar}Parameters of relevance for the magnetosphere of some types of NSs}
  \begin{ruledtabular}
  \begin{tabular}{cccccc}
    & millisecond pulsar & normal pulsar & magnetar & SGR1935+2154\\ \hline
    rotation period $P$(s)
    & $5\times10^{-3}$ & $0.5$ & 5 &3.24\\
    light cylinder radius $R_{\text{LC}}$(cm)
    &$2.386\times10^7$ & $2.386\times10^9$ & $2.386\times10^{10}$ & $1.55\times10^{10}$\\
    magnetic field on the surface $B_{*}$(G)
    & $10^{8}$ & $10^{12}$ & $10^{14}$ & $2.2\times10^{14}$\\
    magnetic field in the outer gap $B$(G)\footnote{We calculate the values under the assumption that the outer gap radius is nearly equivalent to the light cylinder radius.}
    & $7.365\times10^{3}$ & $73.65$ & $7.365$ & $60$\\
    electron number density in the outer gap $n_{\text{e}}$($\text{cm}^{-3}$)\footnote{We use the same assumption.}
  & $1.031\times10^{5}$ & $10.31$ & $0.1031$ & $1.3$\\
  \end{tabular}
  \end{ruledtabular}
\end{table*}

Setting the Landau level $n$ in the initial state, we calculate the true absorption rate as follows:
\begin{align}
  \mathcal{W}_{\text{true-abs}} ^{(\sigma)}=\sum_{n'=n-1} ^{n+1}
  \left(
   \mathcal{W}_{\text{abs}, n\rightarrow n'} ^{(\sigma)}
   -\mathcal{W}_{\text{emis}, n\rightarrow n'} ^{(\sigma)}
  \right)\label{sum_maser}
\end{align}
where we add the contributions from transitions to different Landau levels as indicated by the superscripts ($n\rightarrow n$ means the curvature radiation).
As mentioned earlier, we employ the eigenstates of the spin operator (actually its component parallel to the magnetic field) for the initial and final states.
As we demonstrate later, transitions with a spin flip is negligibly small compared with those without a spin flip.

\subsection{Circular magnetic field\label{Circular magnetic field}}
\begin{figure}[!hbtp]
  \includegraphics[width=86mm]{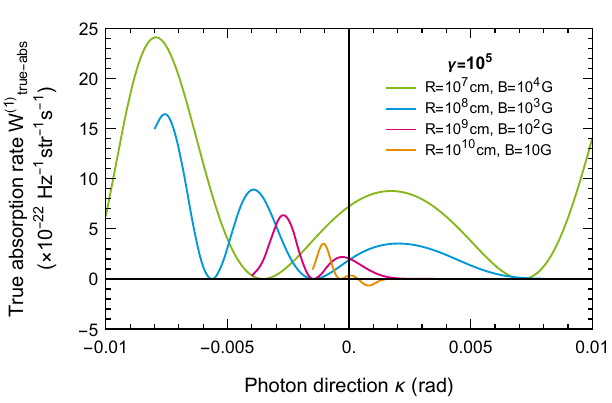}
  \includegraphics[width=86mm]{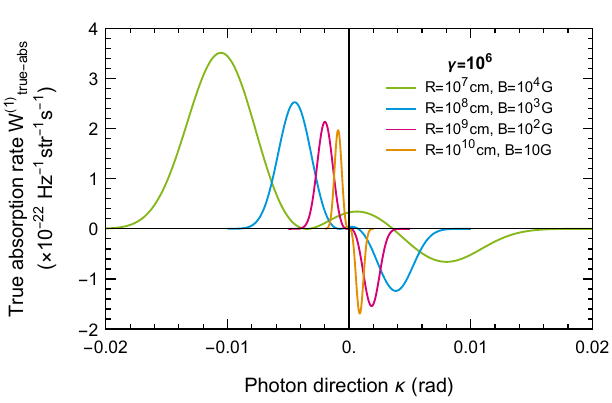}
  \includegraphics[width=86mm]{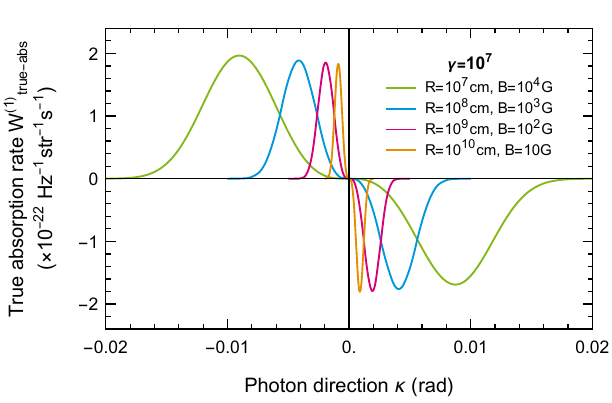}
  \caption{\label{direction dependence}
  The dependence of the true absorption rate on the flight direction of a photon.
  The frequency and polarization of the photon are 1GHz light and $\v{\epsilon}^{(1)}$, respectively.
  The line colors indicate the combination of the values of the radius and magnetic-field strength.
  The initial Landau level is $n=5$.
  The values of the Lorentz factor of the electron are $\gamma=10^5$, $10^6$, and $10^7$ from top to bottom.
  Note that the plotting range for the top panel is different from those in the other two.
  }
\end{figure}

First, we investigate circular magnetic fields ($\alpha=0$).
Tables~\ref{Lorentz5},\ref{Lorentz6},\ref{Lorentz7} show the true absorption rates of a 1GHz photon for different Lorentz factors of the electron.
We vary the radius and field strength according to Table~\ref{Pulsar}.
The flight direction of the photon $\kappa$ is also another important parameter.
In fact, the true absorption rate is sensitive to it and maser occurs in a limited range of $\kappa$ (see Fig.~\ref{direction dependence}).
We hence explore first the $\kappa$-dependence and determine the direction most suitable for maser and adopt it for other discussions.
In this section, we show only the results for polarization $\v{\epsilon}^{(1)}$, since we find essentially no difference between two polarizations.
The initial Landau level is assumed to be $n=5$, which takes account of the synchrotron cooling of electrons \cite{Lu2018, Yang2018}.

Figure~\ref{direction dependence} presents the true absorption rate as a function of $\kappa$ for three different values of $\gamma$.
The colors of the lines indicate the combinations of $R$ and $B$.
Note that maser occurs when the true absorption rate becomes negative.
We can see from the figures that this happens indeed at a range of positive $\kappa$.
The true absorption rates are oscillating, and in the top panel for $\gamma=10^5$, there is a small range of negative true absorption rate only for the combination of $R=10^{10}$cm and $B=10$G (orange line), which may be suitable for magnetars.
As the Lorentz factor gets larger, such a region appears also for other combinations of $R$ and $B$.
The peaks are lowered and the valleys are deepened with increasing $\gamma$ in the right half of the panels in the figures.
At $\gamma=10^7$ (bottom panel), the true absorption rate looks almost symmetric with respect to the origin.
It is also apparent that the valley is closer to $\kappa=0$ for larger $R$.

From Tables~\ref{Lorentz5},\ref{Lorentz6},\ref{Lorentz7}, we find that the direction $\kappa$ most suitable for maser depends only on the radius.
We henceforth fix $\kappa$ to this best value for each radius and investigate the dependences on other parameters.

\begin{table*}
  \caption{\label{Lorentz5}True absorption rate $\mathcal{W}_{\text{true-abs}} ^{(1)}$ (Lorentz factor $\gamma=10^5$)}
  \begin{ruledtabular}
  \begin{tabular}{ccrrrr}
    radius $R$(cm) & strength $B$(G) & \multicolumn{4}{c}{true absorption rate $\mathcal{W}_{\text{true-abs}} ^{(1)}$ ($\text{Hz}^{-1}\text{str}^{-1}\text{s}^{-1}$)}\\
    && \multicolumn{1}{c}{$\kappa=$0.0005rad} & \multicolumn{1}{c}{$\kappa=$0.001rad} & \multicolumn{1}{c}{$\kappa=$0.005rad} & \multicolumn{1}{c}{$\kappa=$0.01rad}\\ \hline
    $10^7$ & $10^{3}$ & $3.432\times10^{-22}$ & $3.381\times10^{-22}$ & $-3.001\times10^{-22}$ & $-5.162\times10^{-22}$\\
           &  $10^{4}$ & $7.996\times10^{-22}$ & $8.500\times10^{-22}$ & $3.789\times10^{-22}$ & $1.407\times10^{-21}$\\
           &  $10^{5}$ & \\ \hline
    $10^8$ & $10^{2}$ & $2.882\times10^{-23}$ & $-5.203\times10^{-23}$ & $-8.783\times10^{-22}$ & $-1.800\times10^{-26}$\\
           & $10^{3}$ & $2.612\times10^{-22}$ & $3.129\times10^{-22}$ & $1.347\times10^{-22}$ & $4.247\times10^{-27}$\\
           & $10^{4}$ & \\ \hline
    $10^9$ & $10$ & $-1.055\times10^{-22}$ & $-6.303\times10^{-22}$ & $-1.227\times10^{-27}$\\
           & $10^{2}$ & $1.356\times10^{-22}$ & $6.684\times10^{-23}$ & $3.355\times10^{-30}$\\
           & $10^{3}$ & \\ \hline
    $10^{10}$ & $1$ & $-7.581\times10^{-22}$ & $-1.548\times10^{-21}$\\
           & $10$ & $-3.005\times10^{-23}$ & $-5.162\times10^{-23}$\\
           & $10^{2}$ & \\
  \end{tabular}
  \end{ruledtabular}
\end{table*}

\begin{table*}
  \caption{\label{Lorentz6}True absorption rate $\mathcal{W}_{\text{true-abs}} ^{(1)}$ (Lorentz factor $\gamma=10^6$)}
  \begin{ruledtabular}
    \begin{tabular}{ccrrrr}
      radius $R$(cm) & strength $B$(G) & \multicolumn{4}{c}{true absorption rate $\mathcal{W}_{\text{true-abs}} ^{(1)}$ ($\text{Hz}^{-1}\text{str}^{-1}\text{s}^{-1}$)}\\
      && \multicolumn{1}{c}{$\kappa=$0.0005rad} & \multicolumn{1}{c}{$\kappa=$0.001rad} & \multicolumn{1}{c}{$\kappa=$0.005rad} & \multicolumn{1}{c}{$\kappa=$0.01rad}\\ \hline
    $10^7$ & $10^{3}$ & $-8.792\times10^{-25}$ & $-9.260\times10^{-24}$ & $-7.578\times10^{-22}$ & $-1.548\times10^{-21}$\\
           &  $10^{4}$ & $3.432\times10^{-23}$ & $3.381\times10^{-23}$ & $-3.001\times10^{-23}$ & $-5.162\times10^{-23}$\\
           &  $10^{5}$ & $7.996\times10^{-23}$ & $8.500\times10^{-23}$ & $3.789\times10^{-23}$ & $1.407\times10^{-22}$\\ \hline
    $10^8$ & $10^{2}$ & $-1.089\times10^{-23}$ & $-7.939\times10^{-23}$ & $-1.424\times10^{-21}$ & $-5.137\times10^{-26}$\\
           & $10^{3}$ & $2.883\times10^{-24}$ & $-5.201\times10^{-24}$ & $-8.783\times10^{-23}$ & $-1.800\times10^{-27}$\\
           & $10^{4}$ & $2.612\times10^{-23}$ & $3.129\times10^{-23}$ & $1.347\times10^{-23}$ & $4.247\times10^{-28}$\\ \hline
    $10^9$ & $10$ & $-9.538\times10^{-23}$ & $-6.337\times10^{-22}$ & $-2.066\times10^{-27}$\\
           & $10^{2}$ & $-1.055\times10^{-23}$ & $-6.303\times10^{-23}$ & $-1.227\times10^{-28}$\\
           & $10^{3}$ & $1.356\times10^{-23}$ & $6.685\times10^{-24}$ & $3.356\times10^{-31}$\\ \hline
    $10^{10}$ & $1$ & $-7.586\times10^{-22}$ & $-1.687\times10^{-21}$\\
           & $10$ & $-7.579\times10^{-23}$ & $-1.548\times10^{-22}$\\
           & $10^{2}$ & $-3.003\times10^{-24}$ & $-5.163\times10^{-24}$\\
  \end{tabular}
  \end{ruledtabular}
\end{table*}

\begin{table*}
  \caption{\label{Lorentz7}True absorption rate $\mathcal{W}_{\text{true-abs}} ^{(1)}$ (Lorentz factor $\gamma=10^7$)}
  \begin{ruledtabular}
    \begin{tabular}{ccrrrr}
      radius $R$(cm) & strength $B$(G) & \multicolumn{4}{c}{true absorption rate $\mathcal{W}_{\text{true-abs}} ^{(1)}$ ($\text{Hz}^{-1}\text{str}^{-1}\text{s}^{-1}$)}\\
      && \multicolumn{1}{c}{$\kappa=$0.0005rad} & \multicolumn{1}{c}{$\kappa=$0.001rad} & \multicolumn{1}{c}{$\kappa=$0.005rad} & \multicolumn{1}{c}{$\kappa=$0.01rad}\\ \hline
    $10^7$ & $10^{3}$ & $-1.023\times10^{-24}$ & $-7.947\times10^{-24}$ & $-7.584\times10^{-22}$ & $-1.687\times10^{-21}$\\
           &  $10^{4}$ & $-8.792\times10^{-26}$ & $-9.260\times10^{-25}$ & $-7.578\times10^{-23}$ & $-1.548\times10^{-22}$\\
           &  $10^{5}$ & $3.432\times10^{-24}$ & $3.381\times10^{-24}$ & $-3.001\times10^{-24}$ & $-5.162\times10^{-24}$\\ \hline
    $10^8$ & $10^{2}$ & $-9.754\times10^{-24}$ & $-7.523\times10^{-23}$ & $-1.489\times10^{-21}$ & $-5.696\times10^{-26}$\\
           & $10^{3}$ & $-1.089\times10^{-24}$ & $-7.939\times10^{-24}$ & $-1.424\times10^{-22}$ & $-5.137\times10^{-27}$\\
           & $10^{4}$ & $2.883\times10^{-25}$ & $-5.201\times10^{-25}$ & $-8.783\times10^{-24}$ & $-1.800\times10^{-28}$\\ \hline
    $10^9$ & $10$ & $-9.308\times10^{-23}$ & $-6.317\times10^{-22}$ & $-2.176\times10^{-27}$\\
           & $10^{2}$ & $-9.538\times10^{-24}$ & $-6.337\times10^{-23}$ & $-2.066\times10^{-28}$\\
           & $10^{3}$ & $-1.055\times10^{-24}$ & $-6.303\times10^{-24}$ & $-1.227\times10^{-29}$\\ \hline
    $10^{10}$ & $1$ & $-7.581\times10^{-22}$ & $-1.702\times10^{-21}$\\
           & $10$ & $-7.586\times10^{-23}$ & $-1.687\times10^{-22}$\\
           & $10^{2}$ & $-7.579\times10^{-24}$ & $-1.548\times10^{-23}$\\
  \end{tabular}
  \end{ruledtabular}
\end{table*}

\begin{figure}[!hbtp]
  \includegraphics[width=86mm]{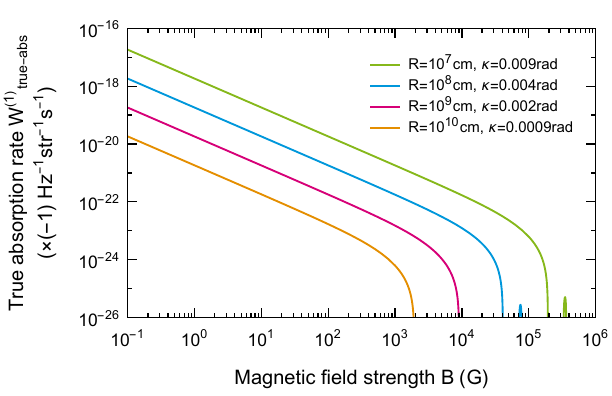}
  \caption{\label{B dependence}
  The dependence of the true absorption rate of a 1GHz photon with the polarization $\v{\epsilon}^{(1)}$ on the magnetic-field strength.
  The flight direction of the photon is fixed to the one, for which the true absorption rate is lowest in Fig.~\ref{direction dependence}.
  The Lorentz factor of the electron is $\gamma=10^7$ and the initial Landau level is $n=5$.
  }
\end{figure}

Figure~\ref{B dependence} exhibits the dependence on the magnetic-field strength.
One finds that the true absorption rate is proportional to $1/B$ and that there may be an upper limit of $B$ for maser, which decreases as the radius becomes larger (see p.~\pageref{B_explanation} for its reason).
It is noted that the typical field strengths for the three types of NSs given in Table~\ref{Pulsar} are below the upper limits.

\begin{figure}[!hbtp]
  \includegraphics[width=86mm]{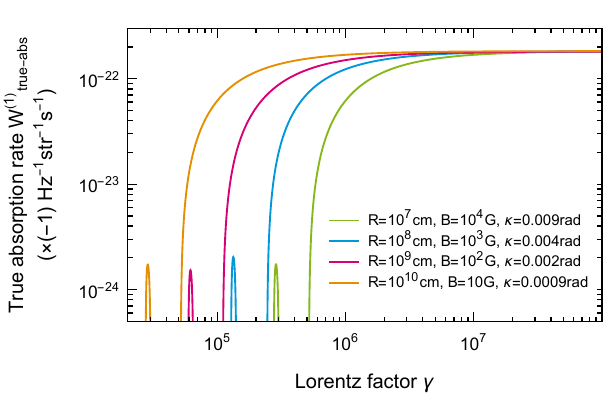}
  \caption{\label{Lorentz dependence}
  The dependence of the true absorption rate of a 1GHz photon with the polarization $\v{\epsilon}^{(1)}$ on the Lorentz factor.
  The line colors denote the combination of the radius, magnetic-field strength and flight direction of the photon. The values of $\kappa$ correspond to the directions, for which the true absorption rate becomes lowest.
  The initial Landau level is $n=5$.
  }
\end{figure}

In Fig.~\ref{Lorentz dependence}, we plot the true absorption rate multiplied with $-1$ as a function of the Lorentz factor again for different combinations of $R$ and $B$.
This time there may be a lower limit, below which maser does not occur.
It is important because the Lorentz factor of secondary particles generated in the cascade of pair productions is smaller than that of the primary particles.
One finds from the figure that the lower limit decreases as the radius gets larger.
It is also observed that the true absorption rates converge to the same value at $\gamma \gtrsim 10^7$.

\begin{figure}[!hbtp]
  \includegraphics[width=86mm]{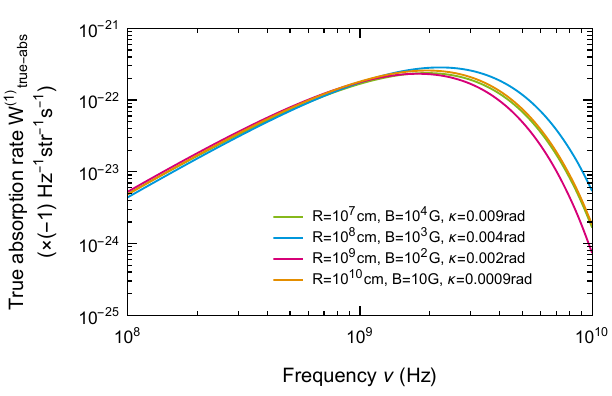}
  \caption{\label{spectrum}
  The dependence of the true absorption rate of a 1GHz photon with the polarization $\v{\epsilon}^{(1)}$ on the photon frequency.
  The line colors specify the combination of the radius, magnetic-field strength and flight direction of the photon.
  The Lorentz factor of the electron is $\gamma=10^7$ and the initial Landau level of the electron is $n=5$.
  }
\end{figure}

So far we have fixed the frequency of the emitted photon to 1GHz.
Figure~\ref{spectrum} presents the frequency dependence.
It is noted that the flight direction of the emitted photon is fixed to the value best suited for maser at 1GHz as mentioned earlier.
It is found that negative absorption rates are realized in wide ranges with a peak around 1GHz, which are insensitive to the combination of $R$ and $B$.

\begin{figure}[!hbtp]
  \includegraphics[width=86mm]{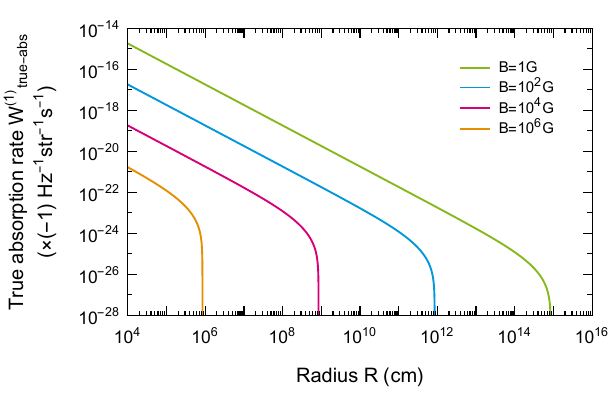}
  \caption{\label{radius dependence}
  The dependence of the true absorption rate of a 1GHz photon with the polarization $\v{\epsilon}^{(1)}$ on the curvature radius.
  The line colors specify the magnetic-field strength.
  The Lorentz factor of the electron is assumed to be $\gamma=10^7$ and the initial Landau level of the electron is set to $n=5$.
  The flight direction of the photon is varied as $\kappa=0.009 \times (R/10^7)^{-1/3}$ because the best direction depends on the radius.
  }
\end{figure}

Figure ~\ref{radius dependence} shows the radius dependence.
So far we have only investigated the true absorption rate for the combinations of $R$ and $B$ based on Table~\ref{Pulsar}.
Here we change $R$ rather arbitrarily, fixing $B$ to different values.
Note that we vary simultaneously the flight direction of the photon as $\kappa=0.009 \times (R/10^7)^{-1/3}$, which corresponds to the best direction for each radius at the fixed values of $B$.
One finds that there is an upper limit of $R$ for maser, which gets smaller as $B$ becomes larger.
It is also observed that the true absorption rate is proportional to $1/R$ except near the upper limit (see p.~\pageref{B_explanation} for its reason).
% This may be estimated from the formulation; the $\theta$-integration is proportional to $1/R$ and the absorption and emission rate are derived from the square of transition current multiplied with some coefficients including $R$ as Eqs.~\eqref{emis} and \eqref{abs}.

\begin{figure*}[!htbp]
  \includegraphics[width=86mm]{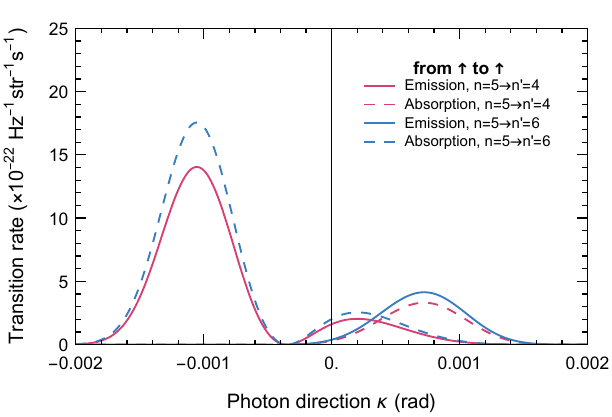}
  \includegraphics[width=86mm]{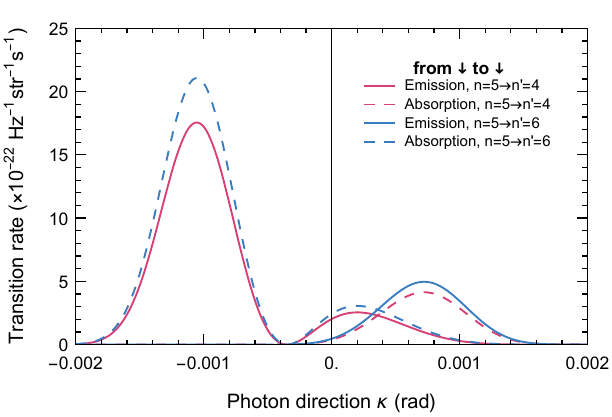}
  \caption{\label{Eachterms_10^10_10^5}
  Same as Fig.~\ref{direction dependence} but for different spin transitions from $\uparrow$ to $\uparrow$ (left panel) and from $\downarrow$ to $\downarrow$ (right panel).
  We set $B=10$G, $R=10^{10}$cm, $\gamma=10^5$, $n=5$.
  The solid lines show the emissivity whereas the dashed lines exhibit the absorptivity.
  Maser occurs in the regions, where a solid line runs above the dashed line.
   % \red{of the same color}
  }
\end{figure*}

We have thus far studied the true absorption rate, summing up the six terms given in Eq.~\eqref{sum_maser}.
We investigate here the individual contributions.
As we mentioned earlier, we employ for the initial and final states the eigenstates of the spin operator projected on the magnetic field.
The dominant transitions are those without a spin flip (see Eq.~\eqref{Coefficients of spin transitions}), i.e., either from $\uparrow$ to $\uparrow$ or from $\downarrow$ to $\downarrow$.
Below we also study the dependence on the spin transition.
Figure~\ref{Eachterms_10^10_10^5} shows the emission and absorption rates for these two spin-transitions separately as a function of the flight direction of the photon, $\kappa$.
One can see from these plots that the individual rates depend on the spin transition, whereas the sum thereof, or the true absorption rate, is essentially the same for the two spin transitions.
Note that we have so far plotted the true absorption rate that are summed over the two spin transitions.

This may be understood from the formulae of the transition rates as follows.
Each of the emission and absorption rates is a sum of four terms as given in Eq.~\eqref{j_mu}.
However, one of them is predominant in fact.
For example, in the case of the spin transition from $\uparrow$ to $\uparrow$, the coefficients in the eigenstates are given for $B=10^4$G, $R=10^{7}$cm, $\alpha=0$rad, $\gamma=10^5$, $n=5$ as
\begin{subequations}
  \begin{align}
    c_1'^{*}c_1-c_3'^{*}c_3&=1,
    \\
    |c_1'^{*}c_2-c_3'^{*}c_4|\approx |c_2'^{*}c_1-c_4'^{*}c_3|&\sim 10^{-10},
    \\
    c_2'^{*}c_2-c_4'^{*}c_4 &\sim 10^{-20}.
  \end{align}
\end{subequations}
There are indeed big differences among the coefficients.
On the other hand, the integrals $I'_1 \sim I'_4$ depend on $\lambda k$ and $B_{pq}$ (see the formulae of each transition), the magnitudes of which can be evaluated as
\begin{align}
  \lambda k &\sim 10^{-7},
\end{align}
from Eq.~\eqref{taylor_order} and
\begin{align}
  |B_{pq}|& \sim |B_{00}|\times (10^{-2})^q. \label{Bpq_estimate}
\end{align}
This approximation in Eq.~\eqref{Bpq_estimate} is derived from the fact that $B_{pq}$ is actually the integration in a limited range around $\theta=0$ because the integrand is violently oscillating.
The integration $|B_{pq}|$ hence gets smaller as the integrand includes a larger number of $\sin{\theta}$ and the numeral $10^{-2}$ in Eq.~\eqref{Bpq_estimate} is obtained from a numerical calculation.
% and $|B_{pq}|$ is $q$ times smaller than the $|B_{00}|$

As a result, $(c_1'^{*}c_1-c_3'^{*}c_3)I'_1$ dominates other contributions in this transition rate.
The other transitions, i.e., from $\downarrow$ to $\downarrow$, can be treated similarly.
Then the results are the followings:
\begin{subequations}
  \begin{align}
    \text{from}\uparrow\text{to}\uparrow&
    \begin{dcases}
      \mathcal{W}_{\text{emis},n\rightarrow n-1} ^{(\sigma)}\approx (n-1) J_1 ^{(\sigma)} \\
      \mathcal{W}_{\text{abs},n\rightarrow n-1} ^{(\sigma)}\approx (n-1) J_2 ^{(\sigma)} \\
      \mathcal{W}_{\text{emis},n\rightarrow n+1} ^{(\sigma)}\approx n J_2 ^{(\sigma)} \\
      \mathcal{W}_{\text{abs},n\rightarrow n+1} ^{(\sigma)}\approx n J_1 ^{(\sigma)} \\
    \end{dcases}\label{esitimate of each terms1}
    ,
    \displaybreak[1]\\
    \text{from}\downarrow\text{to}\downarrow&
    \begin{dcases}
      \mathcal{W}_{\text{emis},n\rightarrow n-1} ^{(\sigma)}\approx n J_1 ^{(\sigma)} \\
      \mathcal{W}_{\text{abs},n\rightarrow n-1} ^{(\sigma)}\approx n J_2 ^{(\sigma)} \\
      \mathcal{W}_{\text{emis},n\rightarrow n+1} ^{(\sigma)}\approx (n+1) J_2 ^{(\sigma)} \\
      \mathcal{W}_{\text{abs},n\rightarrow n+1} ^{(\sigma)}\approx (n+1) J_1 ^{(\sigma)} \\
    \end{dcases}\label{esitimate of each terms2}
    ,
  \end{align}
\end{subequations}
where $J_1 ^{(\sigma)}$, $ J_2 ^{(\sigma)}$ are constants that depend on polarizations of the photon and parameters other than the Landau level.
Using these results, we get the true absorption rate as
\begin{align}
  \mathcal{W}_{\text{true-abs}}^{(\sigma)}=J_1 ^{(\sigma)}-J_2 ^{(\sigma)},
\end{align}
for both transitions.
This is the reason why the true absorption rate is the same although the individual transition rates are different.
In deriving this result, we neglect the transitions between the same Landau level as the emission and absorption rates almost cancel each other.
Note also that the true absorption rate does not depend on the initial Landau level.
Strictly speaking, this is not true, since there is still a minor dependence on it through $H(\theta)$ in $J_1 ^{(\sigma)}$ and $J_2 ^{(\sigma)}$.
However, for the parameter values typical to the NS magnetosphere, the focus in this paper, $H(\theta)$ may be approximated at $\theta \approx 0$, the region dominantly contributing to the integral, as
\begin{align}
  H(\theta) \approx -\rho k \theta.\label{H_approx}
\end{align}
Hence the Landau level does not influence $J_1 ^{(\sigma)}$ and $J_2 ^{(\sigma)}$.
It is again mentioned that the transition rate with a spin flip is negligible as understood from the above evaluations of the integrals $I'_1 \sim I'_4$ and their coefficients.

Although we have investigated the transitions between the excited Landau levels alone, we can also consider those from the ground state similarly.
In this case, the transtion from $\uparrow$ to $\uparrow$ cannot occur according to Eq.~\eqref{esitimate of each terms1}, and the latter two transitions in Eq.~\eqref{esitimate of each terms2} are dominant.
The true absorption rates are consequently equal to those between the excited Landau levels.

Here we point out an interesting scaling.
By the inspection of Tables~\ref{Lorentz5},\ref{Lorentz6},\ref{Lorentz7}, one realizes that the true absorption rates for $\gamma=10^5$, $R=10^7\text{cm}$, $B=10^{3} \text{G}$, $\kappa=0.005 \text{rad}$, and $\gamma=10^6$, $R=10^{7} \text{cm}$, $B=10^{4} \text{G}$, $\kappa=0.005 \text{rad}$ are different only in the exponent.
This can be explained by a similar discussion to the one given above.
From the approximate expression of $H(\theta)$ in Eq.~\eqref{H_approx}, we find that it depends only on $\rho$, and from our numerical calculation, the $\theta$-integration may be proportional to $1/\rho$.
Furthermore, the coefficients in $j^\mu_{\text{abs}} \epsilon_\mu^{(\alpha)}$ scale approximately as
\begin{align}
  (c_1'^{*}c_1-c_3'^{*}c_3) \lambda k
  \propto \frac{1}{\sqrt{B}},
\end{align}
and those in the $j^\mu_{\text{emis}} \epsilon_\mu^{(\alpha)}$ as
\begin{align}
  (c_2'^{*}c_2-c_4'^{*}c_4) \lambda k
  \propto \frac{1}{\sqrt{B}}.
\end{align}
As the emission and absorption rates are proportional to the square of each of them, $\mathcal{W}_{\text{true-abs}} ^{(1)}$ scales as $1/B$ under the assumption of the same radius.\label{B_explanation}
This explains the coincidence we observe in the tables and can also describe Fig.~\ref{B dependence}.

Similarly we can explain the reason why the true absorption rate is proportional to $1/R$ in Fig.~\ref{radius dependence}.
As mentioned above, the $\theta$-integration is proportional to $1/R$.
The emission and absorption rates are derived from the square of transition current multiplied with some coefficients including $R$ as Eqs.~\eqref{emis} and \eqref{abs}.
Hence the true absorption rate is proportional to $(1/R)^2 \times R =1/R$.

Evaluating $H(\theta)$, we can also explain the fact that there is the upper limit to $B$ and the lower limit to $\gamma$ for maser.
The approximation in Eq.~\eqref{H_approx} do not hold in the region where $1/\gamma_2$ in $H(\theta)$ is not negligible.
As $B$ increases or as $\gamma$ decreases, the term $1/\gamma_2$ in Eq.~\eqref{taylor} becomes dominant and $H(\theta)$ can be approximated as
\begin{align}
  H(\theta) \approx -R k \frac{1}{\gamma}\frac{mc^2}{h \nu}\frac{B}{B_c} \Delta n \theta.
\end{align}
Then $e^{i H(\theta)}$ oscillates more violently and the $\theta$-integration tends to be canceled to 0.
This is the reason why $B$ has the upper limit and $\gamma$ has the lower limit for maser.

% Evaluating $H(\theta)$, we can also explain the fact that there is the upper limit \red{to} $B$ and the lower limit \red{to} $\gamma$ for maser as follows: the absolute value of $H(\theta)$ becomes larger either as $B$ increases or as $\gamma$ decreases, since it is proportional to $R B/\gamma$; then \red{$e^{i H(\theta)}$ oscillates more violently and the $\theta$-integration tends to be canceled to 0.}
% This is the reason why $B$ has the upper limit and $\gamma$ has the lower limit for maser.

Note finally that the true absorption rate does not keep increasing as $B$ gets closer to 0 in Fig.~\ref{B dependence}.
This is because, in the weak-field limit of $\lambda k \gg 1$, the expansion of Eq.~\eqref{taylor_before} is no longer valid.
Since maser cannot occur without the magnetic field, we may surmise that the maser emissivity will stop growing at some point and the true absorption rate will eventually become positive again as the magnetic field approaches zero.

\subsection{Helical magnetic field\label{Helical magnetic field}}
\begin{figure*}[!htbp]
  \includegraphics[width=86mm]{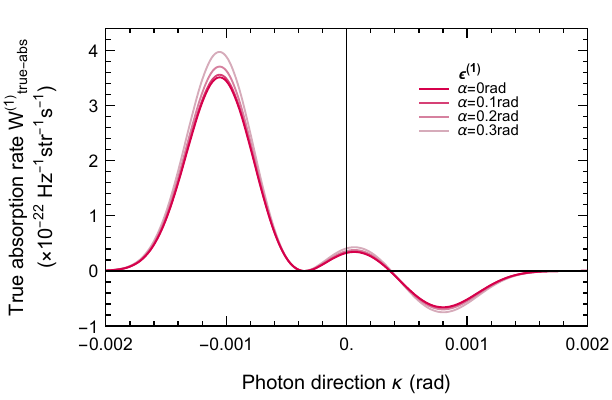}
  \includegraphics[width=86mm]{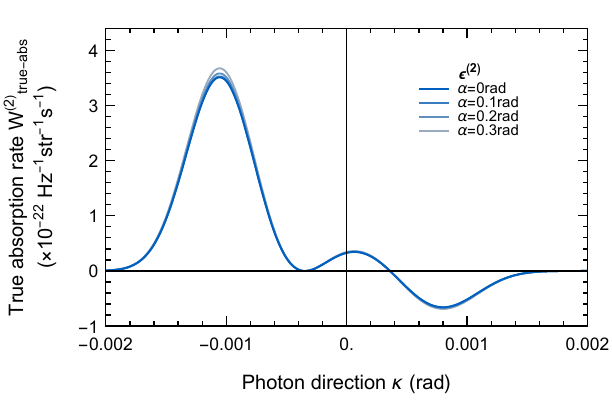}
  \caption{\label{torsion 10^5}
  The true absorption rates of a 1GHz photon with the polarization of $\v{\epsilon}^{(1)}$ (left panel) and $\v{\epsilon}^{(2)}$ (right panel) as a function of the flight direction of the photon for helical magnetic fields.
  We set $B=10$G, $\rho=10^{10}$cm, $\gamma=10^5$, and $n=5$.
  The pitch angle is varied as $\alpha=$0rad, 0.1rad, 0.2rad, 0.3rad.
  }
\end{figure*}

\begin{figure*}[!htbp]
  \includegraphics[width=86mm]{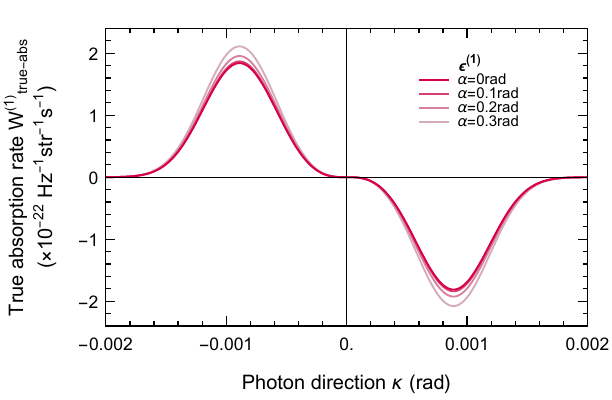}
  \includegraphics[width=86mm]{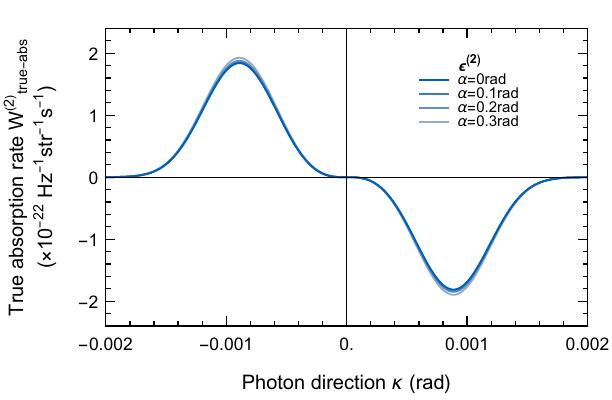}
  \caption{\label{torsion 10^7}
  Same as Fig.~\ref{torsion 10^5} but for $B=10$G, $\rho=10^{10}$cm, $\gamma=10^7$, and $n=5$.
  }
\end{figure*}

In this section, we summarize our findings for the helical magnetic field ($\alpha \neq 0$).
Figures~\ref{torsion 10^5} and \ref{torsion 10^7} show the true absorption rates for different pitch angles as a function of the flight direction of the photon measured from the helical field, $\kappa - \alpha$, for the two polarizations of the photon individually.
Note that in these models we change $R$ with the pitch angle $\alpha$ so that the curvature radius $\rho$ should be fixed.

One finds that the results for the helical field are essentially the same as those for the circular field.
In particular, the photon directions, at which the true absorption rate takes the maximum and minimum values, are unchanged if they are measured from the helical field.
Some differences are found, though.
In fact, the true absorption rates are a little enhanced by the introduction of a non-vanishing pitch angle.
% \blue{
Moreover, we find from the comparison between the left and right panels that the variation produced by the torsion is larger for $\v{\epsilon}^{(1)}$ than for $\v{\epsilon}^{(2)}$.

\subsection{Amplification factor\label{Absorption coefficient estimate}}
In this section, we will roughly estimate to what extent the radiation is coherently amplified by maser.
This may be addressed conveniently with the radiative transfer equation written as
\begin{align}
  \frac{d \mathcal{I}(\v{x},\v{k},t)}{d \ell}&=-\mathcal{A}(\v{k}) \mathcal{I}(\v{x},\v{k},t) +\mathcal{J}(\v{x},\v{k},t)\label{rte},
\end{align}
where $\mathcal{I}(\v{x},\v{k},t)$ is the intensity, or radiation energy per frequency per area per solid angle per time; $\mathcal{A}(\v{k})$ is the net absorption coefficient corrected for the induced emission; $\mathcal{J}(\v{x},\v{k},t)$ is the spontaneous-emission coefficient, which we ignore here; $\ell$ is the distance the light travels.
It is obvious that a negative value of $\mathcal{A}(\v{k})$ implies an exponential amplification of radiation.

\begin{figure}[!hbtp]
  \includegraphics[width=86mm]{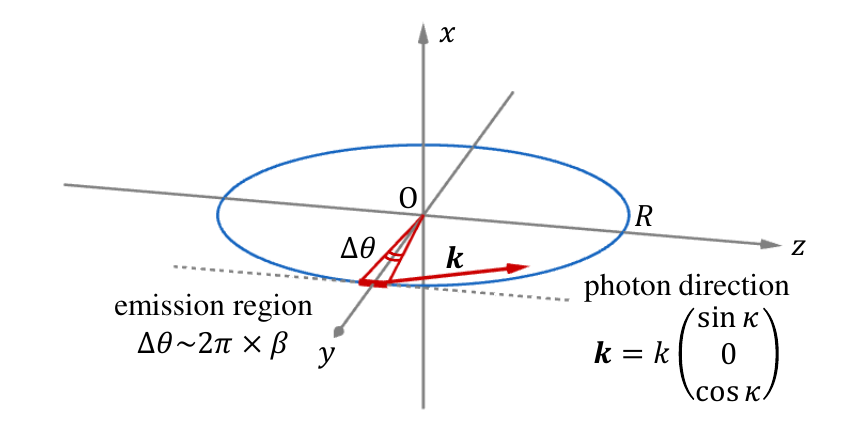}
  \caption{\label{emission_region}Emission region}
\end{figure}

The absorption coefficient $\mathcal{A} ^{(\sigma)}(\v{k})$, which depends on the polarization of the photon, is related to the true absorption rate $\mathcal{W}_{\text{true-abs}} ^{(\sigma)}$ as follows:
\begin{align}
  \mathcal{A} ^{(\sigma)}(\v{k})
  &=\mathcal{W}_{\text{true-abs}} ^{(\sigma)}\times \left( \frac{\nu^2}{c^3}\right)^{-1} \times n_{\text{e}} \times \frac{1}{c} \times \frac{1}{\beta}.
\end{align}
In this equation, since $\mathcal{W}_{\text{true-abs}} ^{(\sigma)}$ contains the density of states, we re-eliminate it by the second term; we multiply the electron number density $n_{\text{e}}$ as $\mathcal{W}_{\text{true-abs}} ^{(\sigma)}$ is the rate for a single electron; the factor $1/c$ is needed to convert the rate per time to the rate per length, and the last term contains all the information on configuration of the magnetic field in the emission region.
In fact, the integration with respect to $\theta$ depends almost solely on the integrand around $\theta=0$ because the $e^{iH(\theta)}$ is oscillating violently.
It means that the radiation occurs in the limited region and we should consider it as the actual emission region (see Fig.~\ref{emission_region}).

We now evaluate the amplification factor when photons are crossing the magnetic field of appropriate configurations.
Assuming that the size of the maser-emission region is of the order of the distance from the NS and taking typical values of $\mathcal{W}_{\text{true-abs}} ^{(\sigma)}=10^{-22} \text{Hz}^{-1}\text{str}^{-1}\text{s}^{-1}$, $\nu=10^9 \text{Hz}$, $\beta=10^{-3}$, we may write the amplification factor as
\begin{align}
  \mathcal{A} ^{(\sigma)}(\v{k})R  &\sim 10^{-17}\text{cm}^{2} \times n_{\text{e}} R.
\end{align}
To explain the very high brightness temperature of $\sim 10^{36}$K \cite{Zhang2020} in FRBs, $\mathcal{A} ^{(\sigma)}(\v{k})R$ needs to be $\gtrsim 10^2$, implying that $n_{\text{e}} R$ should be larger than $\sim 10^{19}$cm$^{-2}$.
For the millisecond pulsar, for example, we have
\begin{align}
  n_{\text{e}} R \sim 10^{12} \text{cm}^{-2} \times \mathcal{M},
\end{align}
where $\mathcal{M}$ is the multiplicity in the cascade of pair production.
One hence finds that $\mathcal{M} \gtrsim 10^7$ is required, which may not be unreasonable.
For other kinds of pulsars, however, the electron number density $n_{\text{e}}$ may not be large enough as the distance to the outer gap from the NS is larger as given in Table~\ref{Pulsar}.

\section{Conclusion\label{Conclusion}}
In this paper we have considered the possibility of the synchro-curvature maser in the quantum regime.
Maser has been one of the major mechanisms advocated for the highly coherent emissions from astronomical objects such as FRBs.
So far the maser in the synchro-curvature radiation, which is expected to occur commonly in non-uniform magnetic fields, has been investigated extensively in classical mechanics alone.
In this paper, we have solved the Dirac equations for highly relativistic electrons in the helical magnetic field with its curvature radius much larger than the Larmor radius of the electron.
Under this condition, the change of magnetic field is adiabatically gradual and we can derive the wave functions of the electron in the helical field from those in the uniform field, using the adiabatic spinor rotations.
We have also formulated the emission and absorption rates based on the perturbation theory for the radiative interactions of the electron.
In so doing, we have adopted for the initial and final electron states the eigenstates of either the spin operator projected onto the magnetic field or the helicity operator.

Then we have numerically evaluated the true absorption rate, which incorporates the contribution from the induced emission, for some parameter values typical to the magnetosphere of NS.
Here the eigenstates of the spin operator projected onto the magnetic field are employed for the initial and final states of the electron.
Maser emissions are characterized by negative values of this true absorption rate.

We have found indeed that there is a range of parameters that produce maser, which are also relevant for FRBs.
To be more specific, we have first demonstrated for the circular field that there is the best direction for maser, which is close to but a bit off the forward direction of the electron motion and depends only on the curvature radius.
Then we have investigated the dependence on other parameters in detail.
We have observed that there is an upper limit for the magnetic-field strength that gives maser, the fact that may favor the outer gap rather than the polar cap as the emission region, whereas there is a lower limit for the Lorentz factor: $\gamma \gtrsim 10^5$.
It has also been demonstrated that maser occurs in a somewhat wide range around $1$GHz.
We have further gone into details, showing that the radiative transitions with a spin flip are negligible compared with those without a flip; and the true absorption rates are essentially the same for the two transitions without a spin flip, i.e., from $\uparrow$ to $\uparrow$ and from $\downarrow$ to $\downarrow$ although the individual contributions from different transitions between the Landau levels do depend on the spin.
Incidentally, we have observed that the latter individual contributions depend not on the initial and final Landau levels separately.
Finally we have demonstrated that the results for the helical magnetic field is essentially the same as for the circular magnetic field although the rather minor changes that the pitch angle introduces are more remarkable for photons with the polarization $\v{\epsilon}^{(1)}$ than for those with $\v{\epsilon}^{(2)}$ for a fixed curvature radius.

Based on these results, we have made a very crude estimate of the amplification factor for the masers that could occur in the outer gap of a millisecond pulsar.
The result suggests that the multiplicity larger than $10^7$ may be required for the very high brightness temperature of FRBs to be obtained.
We have also found that the electron number density may not be large enough for other types of NSs.
These estimates are admittedly very crude, though, and need an improvement based probably on numerical solutions of the radiative transfer equations for more realistic configurations of magnetic field.
It should be also mentioned that the drift motions of an electron in the non-uniform magnetic field have been ignored in this paper and should be the target of future studies.

  \begin{acknowledgments}
    S.Y. is supported by Institute for Advanced Theoretical and Experimental Physics, Waseda University and the Waseda University Grant for Special Research Projects (project number: 2021-C197).
  \end{acknowledgments}
  \begin{widetext}
\appendix
  \section{Helical coordinates\label{A}}
  In this appendix, we introduce the helical coordinates employed in this paper.
  Being deployed along the field line, they are useful for considering the helical magnetic field.

  The helix is expressed in terms of the pitch angle $\alpha$, which is measured from the $yz$ plane, and the radius $R$ of the projected circle as well as the angle $\theta$ from the $y$-axis (see Fig.~\ref{Helical}) as
  \begin{gather}
    \begin{pmatrix}
      x\\
      y\\
      z\\
    \end{pmatrix}
    =
    \begin{pmatrix}
      R\theta\tan{\alpha}\\
      R\cos{\theta}\\
      R\sin{\theta}\\
    \end{pmatrix}.
    \label{helix}
  \end{gather}
  Differentiating this equation with respect to $\theta$, we obtain the tangent vector and its length as
  \begin{gather}
    \begin{pmatrix}
      \dot{x}\\
      \dot{y}\\
      \dot{z}\\
    \end{pmatrix}
    =
    \begin{pmatrix}
      R\tan{\alpha}\\
      -R\sin{\theta}\\
      R\cos{\theta}\\
    \end{pmatrix}
    ,\\
    \sqrt{{\dot{x}}^2+{\dot{y}}^2+{\dot{z}}^2}=\frac{R}{\cos{\alpha}} \equiv d.
  \end{gather}
  Defining the length along the helix $z_{\text{h}}$ as $z_{\text{h}}=d \theta$, we express Eq.~\eqref{helix} as
  \begin{gather}
    \begin{pmatrix}
      x\\
      y\\
      z\\
    \end{pmatrix}
    =
    \begin{pmatrix}
      z_{\text{h}}\sin{\alpha}\\
      R\cos{\left(\tfrac{z_{\text{h}}}{d}\right)}\\
      R\sin{\left(\tfrac{z_{\text{h}}}{d}\right)}\\
    \end{pmatrix}.
  \end{gather}
  The tangent vector $\v{e}_1$, normal vector $\v{e}_2$ and binormal vector $\v{e}_3$ are given, respectively, as
  \begin{gather}
    \v{e}_1=
    \begin{pmatrix}
      \sin{\alpha}\\
      -\cos{\alpha}\sin{\left(\tfrac{z_{\text{h}}}{d}\right)}\\
      \cos{\alpha}\cos{\left(\tfrac{z_{\text{h}}}{d}\right)}\\
    \end{pmatrix}
    ,\quad
    \v{e}_2=
    \begin{pmatrix}
      0\\
      -\cos{\left(\tfrac{z_{\text{h}}}{d}\right)}\\
      -\sin{\left(\tfrac{z_{\text{h}}}{d}\right)}\\
    \end{pmatrix}
    ,\quad
    \v{e}_3=
    \begin{pmatrix}
      \cos{\alpha}\\
      \sin{\alpha}\sin{\left(\tfrac{z_{\text{h}}}{d}\right)}\\
      -\sin{\alpha}\cos{\left(\tfrac{z_{\text{h}}}{d}\right)}\\
    \end{pmatrix}.
  \end{gather}
  The curvature radius $\rho$ is then written as
  \begin{gather}
    \rho=\frac{R}{\cos^2{\alpha}}.
  \end{gather}
  In the neighborhood of the helix, we can deploy the helical coordinates, which are nothing but the cylindrical coordinates locally (see Fig.~\ref{Helical}).
  The Cartesian coordinates are written in terms of the helical coordinates as follows:
  \begin{align}
    \begin{pmatrix}
      x\\
      y\\
      z\\
    \end{pmatrix}
    =
    \begin{pmatrix}
      z_{\text{h}}\sin{\alpha}+r\cos{\phi}\cos{\alpha}
      \\
      (R+r\sin{\phi})\cos{\left(\tfrac{z_{\text{h}}}{d}\right)}+r\cos{\phi}\sin{\alpha}\sin{\left(\tfrac{z_{\text{h}}}{d}\right)}
      \\
      (R+r\sin{\phi})\sin{\left(\tfrac{z_{\text{h}}}{d}\right)}-r\cos{\phi}\sin{\alpha}\sin{\left(\tfrac{z_{\text{h}}}{d}\right)}
      \\
    \end{pmatrix}.
  \end{align}
  The corresponding Jacobian is calculated as
    \begin{align}
      {g}_{ij}=
      \begin{pmatrix}
        1 & 0 & 0\\
        0 & r^2 & 0\\
        0 & 0 & \sin^2{\alpha}+\frac{\cos^2{\alpha}}{R^2}\left(R + r\sin{\phi}\right)^2
        +\frac{r^2}{R^2}\cos^2{\alpha}\sin^2{\alpha}\cos^2{\phi}\\
      \end{pmatrix}.
      \label{Jacobian}
    \end{align}
  The unit vectors aligned with the axes of the helical coordinates are given as
  \begin{align}
    \begin{aligned}
      \v{e}_r
      &=
      \begin{pmatrix}
        \cos{\phi}\cos{\alpha}\\
        \cos{\phi}\sin{\alpha}\sin{\left(\tfrac{z_{\text{h}}}{d}\right)}+\sin{\phi}\cos{\left(\tfrac{z_{\text{h}}}{d}\right)}\\
        -\cos{\phi}\sin{\alpha}\cos{\left(\tfrac{z_{\text{h}}}{d}\right)}+\sin{\phi}\sin{\left(\tfrac{z_{\text{h}}}{d}\right)}\\
      \end{pmatrix},
      \\
      \v{e}_\phi
      &=
      \begin{pmatrix}
        -\sin{\phi}\cos{\alpha}\\
        -\sin{\phi}\sin{\alpha}\sin{\left(\tfrac{z_{\text{h}}}{d}\right)}+\cos{\phi}\cos{\left(\tfrac{z_{\text{h}}}{d}\right)}\\
        \sin{\phi}\sin{\alpha}\cos{\left(\tfrac{z_{\text{h}}}{d}\right)}+\cos{\phi}\sin{\left(\tfrac{z_{\text{h}}}{d}\right)}\\
      \end{pmatrix},
      \\
      \v{e}_{z_\text{h}}
      &=
      \begin{pmatrix}
        \sin{\alpha}\\
        -\cos{\alpha}\sin{\left(\tfrac{z_{\text{h}}}{d}\right)}\\
        \cos{\alpha}\cos{\left(\tfrac{z_{\text{h}}}{d}\right)}\\
      \end{pmatrix}.
    \end{aligned}
  \end{align}
  Although we prefer $z_{\text{h}}$, we also use $\theta$ sometimes for brevity of expressions.

  \section{Laguerre polynomials and the other expressions of the wave functions\label{B}}
  In this appendix, we consider another expression of wave functions, which are also the solutions of Eq.~\eqref{D6}.
  \subsection{Laguerre polynomials}
  We first give the definition of the Laguerre polynomials (Eq.~\eqref{Laguerre}):
  \begin{align}
    L_s ^{(\alpha)} (x) \equiv x^{-\alpha} e^{x} \frac{d^s}{d x^s}
    \left(x^{\alpha+s} e^{-x}\right)
    \quad (s,\alpha=0,1,2\cdots), \label{B-1}
  \end{align}
  which can be written as
  \begin{align}
    L_s ^{(\alpha)} (x)
    =\sum_{k=0} ^{s} (-1)^{s-k} \frac{s! (\alpha+s)!}{k! (s-k)! (\alpha+s-k)!} x^{s-k}.\label{B-2}
  \end{align}
  We have employed it in the main text under the assumption that $s$ and $\alpha$ are both non-negative integers.
  Here we will consider the case of $\alpha<0$.
  Note that there is a limit of differentiation in Eq.~\eqref{B-1} when $\alpha$ is a negative integer.
  Using non-negative integers $\alpha$, we can write $L_s ^{(-\alpha)} (x)$ as
  \begin{align}
    L_s ^{(-\alpha)} (x)=
    \begin{dcases}
      \sum_{k=0} ^{s-\alpha} (-1)^{s-k} \frac{s! (-\alpha+s)!}{k! (s-k)! (-\alpha+s-k)!} x^{s-k}
      & (s-\alpha\geq0)
      \\
      \sum_{k=0} ^{s} (-1)^s \frac{s! (\alpha-s+k-1)!}{k! (s-k)! (\alpha-s-1)!} x^{s-k}
      & (s-\alpha<0)
      \\
    \end{dcases}.\label{B-3}
  \end{align}
  The former case can be derived similarly to Eq.~\eqref{B-2}, but the upper bound of summation is determined by the power of $x^{\alpha+s}$ in Eq.~\eqref{B-1}.
  In the latter case, on the other hand, we need to pay attention to the fact that the power of $x^{\alpha+s}$ in Eq.~\eqref{B-1} is negative.

  From the above equations, we can derive below the following relation:
  \begin{align}
    (-1)^{\alpha} x^{-\alpha} L_{s+\alpha} ^{(-\alpha)}(x)=L_{s} ^{(\alpha)}(x).\label{B-4}
  \end{align}
  % \paragraph{$\alpha\geq0$}
  In fact, for $\alpha \geq 0$, the polynomial $L_{s+\alpha} ^{-\alpha}(x)$ is derived from the upper case of Eq.~\eqref{B-3} as
  \begin{align}
    L_{s+\alpha} ^{(-\alpha)}(x)
    =\sum_{k=0} ^{(s+\alpha)-\alpha}
    (-1)^{(s+\alpha)-k} \frac{(s+\alpha)! (-\alpha+(s+\alpha))!}{k! ((s+\alpha)-k)! (-\alpha+(s+\alpha)-k)!} x^{(s+\alpha)-k}.
  \end{align}
  Hence the left side of Eq.~\eqref{B-4} is calculated as
  \begin{align}
    (-1)^{\alpha} x^{-\alpha} L_{s+\alpha} ^{(-\alpha)}(x)
    =\sum_{k=0} ^{s} (-1)^{s+2\alpha-k}
    \frac{(s+\alpha)! s!}{k! (s+\alpha-k)! (s-k)!} x^{s-k},
  \end{align}
  which is equal to $L_s ^{(\alpha)} (x)$ because of $(-1)^{(s+2\alpha)-k}=(-1)^{s-k}$.
  % \paragraph{$\alpha<0$}
  For $\alpha<0$, on the other hand, the polynomial $L_{s+\alpha} ^{(-\alpha)}(x)$ is written from Eq.~\eqref{B-2} as
  \begin{align}
    L_{s+\alpha} ^{(-\alpha)}(x)
    =\sum_{k=0} ^{s+\alpha}
    (-1)^{(s+\alpha)-k} \frac{(s+\alpha)! (-\alpha+(s+\alpha))!}{k! ((s+\alpha)-k)! (-\alpha+(s+\alpha)-k)!} x^{(s+\alpha)-k}.
  \end{align}
  Then the left side of Eq.~\eqref{B-4} is calculated as
  \begin{align}
    (-1)^{\alpha} x^{-\alpha} L_{s+\alpha} ^{(-\alpha)}(x)
    =\sum_{k=0} ^{s+\alpha}
    (-1)^{(s+2\alpha)-k} \frac{(s+\alpha)! s!}{k! (s+\alpha-k)! (s-k)!} x^{s-k},
  \end{align}
  which is again equal to $L_s ^{(\alpha)} (x)$ according to the upper case of Eq.~\eqref{B-3}.

  \subsection{Wave functions}
  Now we consider the solutions of Eq.~\eqref{D6} that correspond to the case of $\alpha=-l_{\perp \pm}$ in Eqs.~\eqref{D6_1} and \eqref{D6_1-1} as follows:
  \begin{gather}
    \xi_{\pm} (v)=C'_{R_\pm} e^{-\frac{v^2}{2}}v^{-l_{\perp \pm}}L_{s'_\pm}^{\left(-l_{\perp \pm} \right)} \left(v^2 \right),
    \\
    s'_\pm=\frac{1}{4}\left[-l_{\parallel \pm} ^2 \mp 2-2
    +\left(\frac{\lambda}{\hbar c}\right)^2\left(E^2-m^2c^4\right)\right], \nonumber
  \end{gather}
  where $C'_{R_+},C'_{R_-}$ are arbitrary constants.
  Here we define $s'$ as $s'=s'_-$ and use the relation $s'_+=s'_- -1$.
  The solution of Eq.~\eqref{D5} is then given as
  \begin{gather}
    \varphi_R(\v{r})=
    \begin{pmatrix}
      C'_{R_+} e^{i l_{\perp +} \phi}e^{i l_\parallel u}
      e^{-\frac{v^2}{2}}v^{-l_{\perp +}}L_{s'-1}^{\left(-l_{\perp +} \right)} \left(v^2 \right)\\
      C'_{R_-} e^{i l_{\perp -} \phi}e^{i l_\parallel u}
      e^{-\frac{v^2}{2}}v^{-l_{\perp -}}L_{s'}^{\left(-l_{\perp -} \right)} \left(v^2 \right)\\
    \end{pmatrix}
    ,\label{B-11}
    \\
    E_{n,l_\parallel}=\sqrt{m^2 c^4 +\left(\frac{\hbar c}{\lambda}\right)^2\left(4s' +l_\parallel ^2 \right)}
    ,\label{B-12}
  \end{gather}
  which looks different from Eq.~\eqref{D7-1}.
  It can be rewritten by using the formula Eq.~\eqref{B-4}, however, as
  \begin{align}
    \varphi_R(\v{r})
    &=
    \begin{pmatrix}
      C'_{R_+} e^{i l_{\perp +} \phi}e^{i l_\parallel u}
      e^{-\frac{v^2}{2}}v^{-l_{\perp +}}
      (-1)^{-l_{\perp +}}(v^2)^{l_{\perp +}}
      L_{s'-1-l_{\perp +}}^{\left(l_{\perp +} \right)} \left(v^2 \right)\\
      C'_{R_-} e^{i l_{\perp -} \phi}e^{i l_\parallel u}
      e^{-\frac{v^2}{2}}v^{-l_{\perp -}}
      (-1)^{-l_{\perp -}}(v^2)^{l_{\perp -}}
      L_{s'-l_{\perp -}}^{\left(l_{\perp -} \right)} \left(v^2 \right)\\
    \end{pmatrix}
    \\
    &=
    \begin{pmatrix}
      (-1)^{-l_{\perp +}}C'_{R_+} e^{i l_{\perp +} \phi}e^{i l_\parallel u}
      e^{-\frac{v^2}{2}}v^{l_{\perp +}}
      L_{s'-1-l_{\perp +}}^{\left(l_{\perp +} \right)} \left(v^2 \right)\\
      (-1)^{-l_{\perp -}}C'_{R_-} e^{i l_{\perp -} \phi}e^{i l_\parallel u}
      e^{-\frac{v^2}{2}}v^{l_{\perp -}}
      L_{s'-l_{\perp -}}^{\left(l_{\perp -} \right)} \left(v^2 \right)\\
    \end{pmatrix}.
  \end{align}
  Here $l_{\perp \pm}$ should satisfy $l_{\perp +}=l_{\perp -}-1$ for the same reason given in Sec.~\ref{Dirac equation of a charged particle_in a static magnetic field}.
  We define $l_{\perp}$ as $l_{\perp}=l_{\perp -}$, change notations $s'\rightarrow n$, $s'-l_{\perp}\rightarrow s$, and finally obtain
  \begin{align}
    \varphi_R(\v{r})
    &=
    \begin{pmatrix}
      (-1)^{-l_{\perp}+1}C'_{R_+} e^{i (l_{\perp}-1) \phi}e^{i l_\parallel u}
      e^{-\frac{v^2}{2}}v^{l_{\perp}-1}
      L_{s}^{\left(l_{\perp}-1 \right)} \left(v^2 \right)\\
      (-1)^{-l_{\perp}}C'_{R_-} e^{i l_{\perp} \phi}e^{i l_\parallel u}
      e^{-\frac{v^2}{2}}v   ^{l_{\perp}}
      L_{s}^{\left(l_{\perp} \right)} \left(v^2 \right)\\
    \end{pmatrix},
  \end{align}
  which is equivalent to the Eq.~\eqref{D7-1}.
  This justifies our use of the expression in Sec.~\ref{Dirac equation of a charged particle_in a static magnetic field}.
  Note that $l_{\parallel}$ is a real number, whereas $n$ and $s$ are non-negative integers, and $l_{\perp}$ is an integer that satisfies $l_{\perp}\leq n$.

  \section{\texorpdfstring{Calculation of $\Delta l_{\parallel}$}{Calculation of Delta l_parallel}\label{C}}
  From the energy conservation, we obtain
  \begin{align}
    h \nu=\frac{\hbar c}{\lambda}
    \left[
    \Delta l_\parallel
    +\left(1+2\frac{B}{B_c}n\right)\left(\frac{l_{\parallel}}{2\gamma^2}-\frac{l'_{\parallel}}{2\gamma '^2}\right)
    +2\frac{B}{B_c}\Delta n\cdot \frac{l'_{\parallel}}{2\gamma '^2}
    \right].\label{delta l}
  \end{align}
  % When the electron's energy $E$ is much larger than the photon's energy $h\nu$, which we always assume to be true in this paper, we obtain the following approximate expressions:
  % \begin{align}
  %   \frac{l'_{\parallel}}{2\gamma '^2}
  %   &=\frac{l'_{\parallel}}{2(E'/mc^2)^2}
  %   =\frac{l'_{\parallel}}{2\gamma^2}\left(1-\frac{h \nu}{E}\right)^{-2},
  % \end{align}
  % which gives
  % \begin{align}
  %   \frac{l_{\parallel}}{2\gamma^2}-\frac{l'_{\parallel}}{2\gamma '^2}
  %   &=\frac{1}{2\gamma^2}\left(l_{\parallel}-(l_{\parallel}-\Delta l_\parallel)\left(1-\frac{h \nu}{E}\right)^{-2}\right).
  % \end{align}
  Using the following relation:
  \begin{align}
      \frac{l'_{\parallel}}{2\gamma '^2}
      &=\frac{l'_{\parallel}}{2(E_{n',l_\parallel'}/mc^2)^2}
      =\frac{l_{\parallel}-\Delta l_{\parallel}}{2\gamma^2}\left(1-\frac{h \nu}{E_{n,l_\parallel}}\right)^{-2},
  \end{align}
  Eq.~\eqref{delta l} can be rewritten approximately as
  \begin{align}
    h \nu& = \frac{\hbar c}{\lambda} \Delta l_\parallel
    \left[
    1+\frac{1}{2\gamma^2}
    \left(1-\frac{h \nu}{E_{n,l_\parallel}}\right)^{-2}
    \left(1+2\frac{B}{B_c}n -2\frac{B}{B_c}\Delta n\right)
    \right] \nonumber
    \\
    &\hspace{30mm}
    +\frac{\hbar c}{\lambda} l_{\parallel}\cdot \frac{1}{2\gamma^2}
    \left[
    1+2\frac{B}{B_c}n
    -\left(1-\frac{h \nu}{E_{n,l_\parallel}}\right)^{-2}
    \left(1+2\frac{B}{B_c}n-2\frac{B}{B_c}\Delta n\right)
    \right].\label{B4}
  \end{align}
  Defining
  \begin{align}
    X&\equiv
    \left(1-\frac{h \nu}{E_{n,l_\parallel}}\right)^{-2}
    \left(1+2\frac{B}{B_c}n -2\frac{B}{B_c}\Delta n\right),
    \displaybreak[1] \\
    Y&\equiv
    1+2\frac{B}{B_c}n
    -\left(1-\frac{h \nu}{E_{n,l_\parallel}}\right)^{-2}
    \left(1+2\frac{B}{B_c}n-2\frac{B}{B_c}\Delta n\right),
    \end{align}
  we simplify Eq.~\eqref{B4} as
  \begin{align}
    h \nu&=\frac{\hbar c}{\lambda} \Delta l_\parallel
    \left(1+\frac{1}{2\gamma^2} X \right)
    +\frac{\hbar c}{\lambda} l_{\parallel}\cdot \frac{1}{2\gamma^2} Y,
  \end{align}
  which can be solved as
  \begin{align}
    \Delta l_\parallel& \approx \frac{\lambda \omega}{c}
    \left(1-\frac{E_{n,l_\parallel}}{h \nu}\cdot \frac{1}{2\gamma^2}Y\right)
    \left(1-\frac{1}{2\gamma^2}X\right)
    \displaybreak[1] \\
    &\approx
    \lambda k
    \left(1-\frac{1}{2\gamma^2}\left(X+\frac{E_{n,l_\parallel}}{h \nu}Y\right)
    +\mathcal{O}\left(\frac{1}{\gamma^4}\right)\right).
  \end{align}

  We can calculate $X+(E_{n,l_\parallel}/h \nu)Y $ as
  \begin{align}
    X+\frac{E_{n,l_\parallel}}{h \nu}Y
    &=
    \left(1-\frac{h \nu}{E_{n,l_\parallel}}\right)^{-2}
    \left(1+2\frac{B}{B_c}n -2\frac{B}{B_c}\Delta n\right)\nonumber\\
    &\hspace{30mm}+\frac{E_{n,l_\parallel}}{h \nu}
    \left[
    1+2\frac{B}{B_c}n
    -\left(1-\frac{h \nu}{E_{n,l_\parallel}}\right)^{-2}
    \left(1+2\frac{B}{B_c}n-2\frac{B}{B_c}\Delta n\right)
    \right]
    \displaybreak[1] \\
    &\approx
    -\left(1+\frac{h \nu}{E_{n,l_\parallel}}\right)
    \left(1+2\frac{B}{B_c}(n-\Delta n)\right)
    +
    2\frac{E_{n,l_\parallel}}{h \nu}\frac{B}{B_c}\Delta n
    +\mathcal{O}\left(\left(\frac{h \nu}{E_{n,l_\parallel}}\right)^2\right).
  \end{align}
  \\
  Finally we derive $\Delta l_\parallel$ as
  \begin{gather}
    \Delta l_\parallel=\lambda k
    \left(
    1+\frac{1}{2\gamma_2}
    \right),
    \\
    \frac{1}{\gamma_2}
    \equiv
    \frac{1}{\gamma^2}\left[
    \left(1+\frac{h\nu}{E_{n,l_\parallel}}\right)
    \left(1+2\frac{B}{B_{\text{c}}}n'\right)
    -
    2\frac{E_{n,l_\parallel}}{h \nu}\frac{B}{B_{\text{c}}}\Delta n
    \right].
  \end{gather}
  \end{widetext}

% \nocite{*}
\bibliography{Synchro_curvature_maser_arXiv}
\end{document}